\documentclass{jpsj2}
\newcommand{\bq}{\boldsymbol{q}}

\newcommand{\bm}{\boldsymbol{m}}

\newcommand{\bR}{\boldsymbol{R}}
\newcommand{\bQ}{\boldsymbol{Q}}
\newcommand{\be}{\boldsymbol{e}}

\title{Numerical Study of Multiple Helical Spin Density Waves and Vortex Spin Structures in Itinerant Electron System}

\author{Yoshiro Kakehashi\thanks{yok@sci.u-ryukyu.ac.jp, to be published in J. Phys. Soc. Jpn.}}
\inst{Department of Physics and Earth Sciences, Faculty of Science,\\
University of the Ryukyus, \\
1 Senbaru, Nishihara, Okinawa, 903-0213, Japan} 

\abst{
Multiple spin density waves and related vortex-type magnetic structures in the itinerant electron system with inversion symmetry have been investigated on the basis of the Hubbard model and the generalized Hartree-Fock approximation combined with the recursion method.  Possible magnetic phase diagram on the fcc lattice is presented in the space of the Coulomb interaction strength and the electron occupation number.  It is found that the 2$Q$ multiple helical spin density waves (SDW) with half-skyrmion half-antiskyrmion paired structure, the 3$Q$ multiple helical SDW with 3 dimensional type vortex structure, the 4$Q$ multiple SDW with partially ordered state, and a new type of 12$Q$ multiple SDW showing both the half-skyrmion-type vortex structure and the partially ordered state can be stabilized around the Stoner instability line.  The stability of their structures is explained on the basis of their electronic structures. 
}

\kword{ itinerant magnetism, multiple helical spin density waves, vortex spin structure,  magnetic skyrmions, antiferromagnetic skyrmions, half-skyrmion, Hubbard model, 
generalized Hartree-Fock approximation }

\begin{document}
\maketitle

\section{Introduction}

Magnetic skyrmions showing a vortex-like magnetic structure~\cite{skyr61,bogd89,bogd94,bogd01,ross06} have recently been found in the B20-type transition metal compounds such as MnSi~\cite{muhl09} and FeGe~\cite{yu11}.  Since they have a new type of topologically protected particle property and have potential applications to magnetic devices, their basic and dynamical properties have been much investigated in the last decade~\cite{fert17,ever18}.  Exploring experimental efforts have further led to the discoveries of new kinds of magnetic skyrmion structures in the past few years~\cite{yu18,zhan16,fuji19}.  The half-skyrmion type structure described by the 2$Q$ Multiple Helical Spin Density Waves (2$Q$-MHSDW) have been found in the thin films of the $\beta$-Mn-type Co${}_{8}$Zn${}_{9}$Mn${}_{3}$ compound~\cite{yu18}, and the 3 dimensional type skyrmions described by the 3$Q$-MHSDW have been reported in the MnGe compound~\cite{zhan16}.  Furthermore it has been found that the MnSi${}_{1-x}$Ge${}_{x}$ ($0.25 < x < 0.7$) alloys form a new kind of  4$Q$ Multiple Spin Density Waves (4$Q$-MSDW)~\cite{fuji19}.
These experimental results indicate that there exist a variety of Multiple Spin Density Waves (MSDW) and related vortex-type spin structures in itinerant electron system.

Since the compounds mentioned above lack the inversion symmetry, the magnetic skyrmions have been explained first by means of a competition between the ferromagnetic interaction and the Dzyaloshinski-Moriya (DM) interaction~\cite{dzya58,mori60} under magnetic field~\cite{bak80,yi09,yu10}.
Later, Okubo {\it et al.} found that the competing magnetic interactions can also stabilize the magnetic skyrmions even in the system with inversion symmetry~\cite{okubo12}.  Lin and Hayami reported that the competing interactions with easy-axis anisotropy can stabilize skyrmions~\cite{lin16}.  Most of the theoretical works, however, are based on the localized model or its continuum version, so that they are not applicable to the magnetic compounds with the MSDW and vortex-type magnetic structures mentioned above, because they are the  itinerant electron systems in which magnetic electrons hop from site to site and form the $d$ bands~\cite{hub63,hub64,gutz63,gutz64,kana63,moriya85,kake13}. The microscopic origin of the appearance of these magnetic structures and their electronic structure have not yet been understood.

We have recently clarified with the help of the Application Visualization System (AVS) image analysis that the MHSDW describe various vortex-type magnetic structures including skyrmion structures~\cite{kake18}.  Furthermore we have shown on the basis of the Ginzburg-Landau (GL) theory that the vortex type of magnetic structures can be stabilized even in the itinerant electron system with inversion symmetry, and suggested that an antiferromagnetic skyrmion structure is possible on the fcc lattice near the half-filled electron number, using the Hubbard model~\cite{hub63,hub64,gutz63,gutz64,kana63} and the Generalized Hartree-Fock (GHF) approximation.  It has not yet been clarified, however, whether or not the other MHSDW and related vortex structures appear in the whole space of the Coulomb interaction energy parameter $U$ and the electron number per site $n$, and the associated magnetic phase diagram on the fcc lattice has not yet been established even on the level of the mean-field approximation.

In order to clarify the stability of the new-type of MSDW and vortex magnetic structures in the itinerant electron system with inversion symmetry, we performed systematic GHF calculations for the Hubbard model on the fcc lattice at zero temperature varying the Coulomb interaction $U$ and the electron number $n$.  
In this paper, we present the magnetic structures and the magnetic phase diagram of the Hubbard model on the fcc lattice obtained in our numerical calculations.
We will show that a variety of the MSDW and vortex-type magnetic structures are possible even in the itinerant electron system with inversion symmetry under zero magnetic field.

We found in our numerical calculations the 2$Q$ Multiple Helical SDW consisting of the half-skyrmion and half-antiskyrmion vortex pairs (2QH), the 3$Q$ Multiple Helical SDW with 3 dimensional twisted vortex structure (3QH), the antiferromagnetic (AF) type 3QH showing the AF-base vortex structure, and the 4$Q$ Multiple SDW (4QMSDW) with a partially ordered state. Furthermore, we found a new type of vortex magnetic structure consisting of the 12$Q$ Multiple SDW (12QMSDW) which shows both the half-skyrmion type vortex structure and partially ordered state.

In the following section, we briefly explain our numerical method to calculate magnetic structures of the MHSDW for the Hubbard model.  In Sect. 3.1, we explain the input magnetic structures in the self-consistent calculations.  In Sect. 3.2, we present the magnetic phase diagram obtained by the present approach.  Because of the GHF self-consistent equations to be satisfied, the output magnetic structures may be different from the input ones.  We obtain therefore various magnetic structures; the self-consistent 2QH structure, the self-consistent 3QH, the self-consistent AF-base 3QH, the self-consistent 4QMSDW, and the 12QMSDW.  We present these structures and discuss their stability in details in Sect. 3.3 $\sim$ 3.6, as well as the other MSDW in Sect. 3.7.  We summarize the results obtained in the present work in Sect. 4.

\section{Method of Calculations for Multiple Spin Density Waves}

     We adopt the Hubbard model~\cite{hub63,hub64} to investigate the stability of the vortex type magnetic structures in itinerant electron system with inversion symmetry. 
\begin{eqnarray}
H = \sum_{i,\sigma} \epsilon_{0} n_{i\sigma} 
+ \sum_{i, j, \sigma} t_{i j} a_{i \sigma}^{\dagger} a_{j \sigma}
+ \sum_{i} U n_{i \uparrow}n_{i \downarrow} \ .
\label{hub-h}
\end{eqnarray}
Here $\epsilon_{0}$,  $t_{i j}$, and $U$ are the atomic level, the nearest-neighbor transfer integral $t$ between sites $i$ and $j$, and the intra-atomic Coulomb interaction energy parameter, respectively.  
$a_{i \sigma}^{\dagger} (a_{i \sigma})$ is the creation 
(annihilation) operator for an electron with spin $\sigma$ on site $i$,
and $n_{i\sigma}=a_{i \sigma}^{\dagger} a_{i \sigma}$ is the number
operator on site $i$ for spin $\sigma$. 

In order to describe the noncollinear magnetic structure in the simplest way, we apply the Generalized Hartree-Fock (GHF) approximation.  In this approximation, we introduce locally rotated coordinates on each site, and adopt the Hartree-Fock approximation.  We then obtain the GHF Hamiltonian as follows.
\begin{eqnarray}
H = \sum_{i \alpha j \gamma} a_{i \alpha}^{\dagger} H_{i \alpha j \gamma} a_{j \gamma}
- \sum_{i} \frac{1}{4} U ( \langle n_{i} \rangle^2 - \langle \boldsymbol{m}_{i} \rangle^2 ) \ .
\label{ghf-h}
\end{eqnarray}
The one electron Hamiltonian matrix element $H_{i \alpha j \gamma}$ is given by
\begin{eqnarray}
H_{i \alpha j \gamma} = 
\big[ ( \epsilon_{0} + \frac{1}{2} U \langle n_{i} \rangle ) \delta_{\alpha \gamma} - 
\frac{1}{2} U \langle \boldsymbol{m}_{i} \rangle \cdot (\boldsymbol{\sigma})_{\alpha \gamma}
\big] \delta_{ij} + t_{ij} \delta_{\alpha\gamma} (1-\delta_{ij}) .
\label{hmatrix}
\end{eqnarray}
Here $\boldsymbol{\sigma}$ are the Pauli spin matrices.  $\langle n_{i} \rangle$ and 
$\langle \boldsymbol{m}_{i} \rangle$ are the average local charge and magnetic moment on site $i$, respectively.

The local charge and magnetic moment in Eqs. (\ref{ghf-h}) and (\ref{hmatrix}) are given in the GHF as
\begin{eqnarray}
\langle n_{i} \rangle = \int d\omega f(\omega) \sum_{\alpha}
\rho_{i\alpha\alpha}(\omega) ,
\label{chargeni}
\end{eqnarray}
\begin{eqnarray}
\langle \boldsymbol{m}_{i} \rangle = 
\sum_{\alpha\gamma} (\boldsymbol{\sigma})_{\alpha\gamma} 
\int d\omega f(\omega)
\rho_{i\gamma\alpha}(\omega) .
\label{spinmi}
\end{eqnarray}
Here $f(\omega)$ is the Fermi distribution function at zero temperature.
$\rho_{i\alpha\gamma}(\omega)$ is the local density of states (DOS) for one electron Hamiltonian (\ref{hmatrix}).
\begin{eqnarray}
\rho_{i\alpha\gamma}(\omega) = \sum_{k} \langle i | k \rangle_{\alpha} 
\delta(\omega -\epsilon_{k}+\mu) \langle k | i \rangle_{\gamma} \ , 
\label{rhoag}
\end{eqnarray}
where $\mu$ is the Fermi energy, $\epsilon_{k}$ and $\langle i | k \rangle_{\alpha}$
are eigen value and eigen vector for Hamiltonian  (\ref{hmatrix}), respectively.
These DOS are given by the on-site one electron Green function  
$G_{i\alpha i\gamma}(z) = [(z-H)^{-1}]_{i\alpha i\gamma}$ as follows.
\begin{eqnarray}
\rho_{i\alpha\gamma}(\omega) = \frac{(-)}{\pi} {\rm Im} \, G_{i\alpha i\gamma}(z) \ .
\label{rhogag}
\end{eqnarray}
Here $z = \omega + i\delta$, $\delta$ being a positive definite infinitesimal number.

Equations  (\ref{hmatrix}),  (\ref{chargeni}),  (\ref{spinmi}), and  (\ref{rhogag}) form the self-consistent equations to obtain the local charges and magnetic moments, $\{ \langle n_{i} \rangle \}$ and $\{ \langle \boldsymbol{m}_{i} \rangle \}$.
The ground-state energy is given by 
\begin{eqnarray}
E = \mu N + \int d\omega f(\omega) \omega \rho(\omega) 
- \sum_{i} \frac{1}{4} U ( \langle n_{i} \rangle^2 - \langle \boldsymbol{m}_{i} \rangle^2 ) \ .
\label{gener}
\end{eqnarray}
Here $N$ is the total electron number and $\rho(\omega)$ is the total DOS given by $\rho(\omega)=\sum_{i\alpha} \rho_{i\alpha\alpha}(\omega)$.

In order to solve the self-consistent equations for arbitrary magnetic structure, we calculated the Green function using the recursion method~\cite{hay75,heine80}.
In this method, we transform the Hamiltonian (\ref{hmatrix}) into a tridiagonal matrix using  a recursive unitary transformation.  The diagonal Green function $G_{i\alpha i\alpha}(z)$, for example, is then expressed by a continued fraction as follows.
\begin{eqnarray}
G_{i\alpha i\alpha}(z) = 
\cfrac{1}{z - a_{1} - 
\cfrac{|b_{1}|^{2}}
{z - a_{2} -
\cfrac{|b_{2}|^{2}} 
{\ldots \cfrac{\ddots}
{\ldots
- \cfrac{|b_{n-1}|^{2}}
{z - a_{n} - T_{n}(z)} 
}}}} \ .
\label{gcont}
\end{eqnarray}
We obtain the recursion coefficients $\{ a_{m}, b_{m} \}$ up to the $n$-th numerically from the Hamiltonian matrix elements using the recursion algorithm.
We approximate the higher-order coefficients by their asymptotic values $a_{\infty}$, $b_{\infty}$, so that we obtain an approximate form of the terminator $T_{n}(z)$ as follows.
\begin{eqnarray}
T_{n} \approx T_{\infty} = \frac{1}{2} \Big(
z - a_{\infty} - \sqrt{(z - a_{\infty})^{2} - 4 |b_{\infty}|^{2}}
\Big) \ .
\label{tinfty}
\end{eqnarray}

For the calculations of $\langle m_{ix} \rangle$ ($\langle m_{iy} \rangle$), we need off-diagonal Green functions $G_{i\uparrow i\downarrow}+G_{i\downarrow i\uparrow}$ ($i (G_{i\uparrow i\downarrow}-G_{i\downarrow i\uparrow})$).  In this case, we introduce the basis set $|i1 \rangle$, $|i2 \rangle$ ($|i3 \rangle$, $|i4 \rangle$) which diagonalize 
$\sigma_{x}$ ($\sigma_{y}$), and calculate 
$G_{i\uparrow i\downarrow}+G_{i\downarrow i\uparrow}$ 
($i (G_{i\uparrow i\downarrow}-G_{i\downarrow i\uparrow})$) using the diagonal matrices 
$G_{i1 i1}$ and $G_{i2 i2}$ ($G_{i3 i3}$ and $G_{i4 i4}$) in the same way~\cite{kake13}.

In the present calculations, we considered the Hubbard model on the fcc lattice, and 
made a large cubic cluster consisting of $10 \times 10 \times 10$ fcc unit cells on a computer, which is further surrounded by 26 cubic clusters with the same size and the same magnetic structure.  We obtained the recursion coefficients $\{ a_{m}, b_{m} \}$ up to the 10th level for 4000 atoms in the central $10 \times 10 \times 10$ cluster, and solved the self-consistent equations (\ref{chargeni}) and (\ref{spinmi}).

\section{Numerical Results}

\subsection{Input magnetic structures}

The Skyrmion-type vortex structures are expressed by the multiple helical spin density waves (MHSDW) as discussed in our last paper~\cite{kake18}.  We considered in the present calculations the 2$Q$-MHSDW (2QH) and 3$Q$-MHSDW (3QH) on the fcc lattice.
Their local magnetic moments (LM) on site $l$ are expressed as follows. 
\begin{align}
\langle \bm_{l} \rangle &= m \, \big( 
\boldsymbol{j} \cos \boldsymbol{Q}_{1} \! \cdot \! \boldsymbol{R}_{l} +
\boldsymbol{k} \sin \boldsymbol{Q}_{1} \! \cdot \! \boldsymbol{R}_{l} +
\boldsymbol{k} \cos \boldsymbol{Q}_{2} \! \cdot \! \boldsymbol{R}_{l} +
\boldsymbol{i} \sin \boldsymbol{Q}_{2} \! \cdot \! \boldsymbol{R}_{l} \big) \ ,
\label{2qm}
\end{align}
\begin{align}
\langle \bm_{l} \rangle &= m \, \big(  
\boldsymbol{j} \cos \boldsymbol{Q}_{1} \! \cdot \! \boldsymbol{R}_{l} +
\boldsymbol{k} \sin \boldsymbol{Q}_{1} \! \cdot \! \boldsymbol{R}_{l} +
\boldsymbol{k} \cos \boldsymbol{Q}_{2} \! \cdot \! \boldsymbol{R}_{l} +
\boldsymbol{i} \sin \boldsymbol{Q}_{2} \! \cdot \! \boldsymbol{R}_{l} \nonumber \\
&\ \ \ \ +
\boldsymbol{i} \cos \boldsymbol{Q}_{3} \! \cdot \! \boldsymbol{R}_{l} +
\boldsymbol{j} \sin \boldsymbol{Q}_{3} \! \cdot \! \boldsymbol{R}_{l} \big) \ .
\label{3qm}
\end{align}
Here $\bR_{l}$ denotes the position vector of magnetic moment $\langle \bm_{l} \rangle$.  $\boldsymbol{i}$, $\boldsymbol{j}$, and $\boldsymbol{k}$ are the unit vectors along the $x$, $y$, and $z$ axes, respectively.  The wave vectors $\bQ_{1}$, $\bQ_{2}$, and $\bQ_{3}$ are defined by $\bQ_{1}=(q,0,0)$, $\bQ_{2}=(0,q,0)$, and $\bQ_{3}=(0,0,q)$, respectively.  $q$ is a wave number in unit of $2\pi/a$, $a$ being the lattice constant of the fcc cubic unit cell.
These $\bQ$ vectors are perpendicular to the rotational planes of each helical wave.
%
%
\begin{figure}[htbp]
\begin{center}
\begin{minipage}{4cm}
\begin{center}
\vspace*{3cm}
\includegraphics[width=4cm]{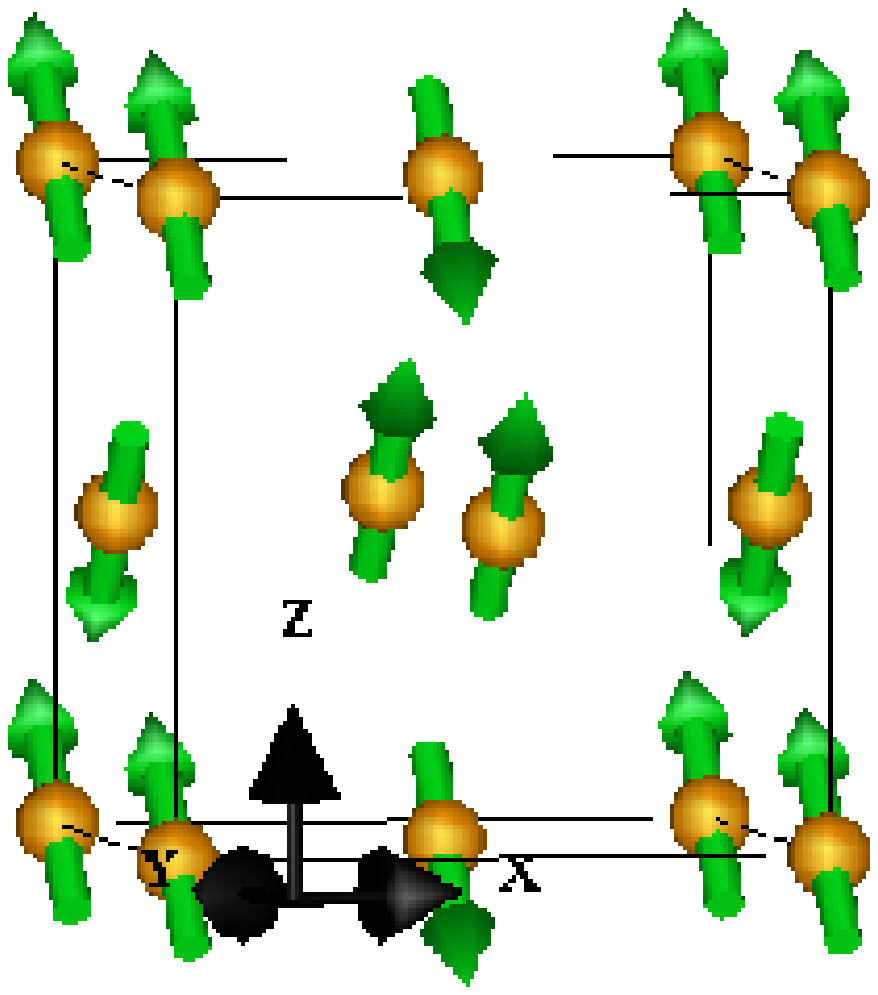} 
\vspace{-0.5cm} \\
(a) 
\vspace{5mm} \\
\end{center}
\end{minipage}
\hspace{0.7cm}
\begin{minipage}{10cm}
\begin{center}
\includegraphics[width=10cm]{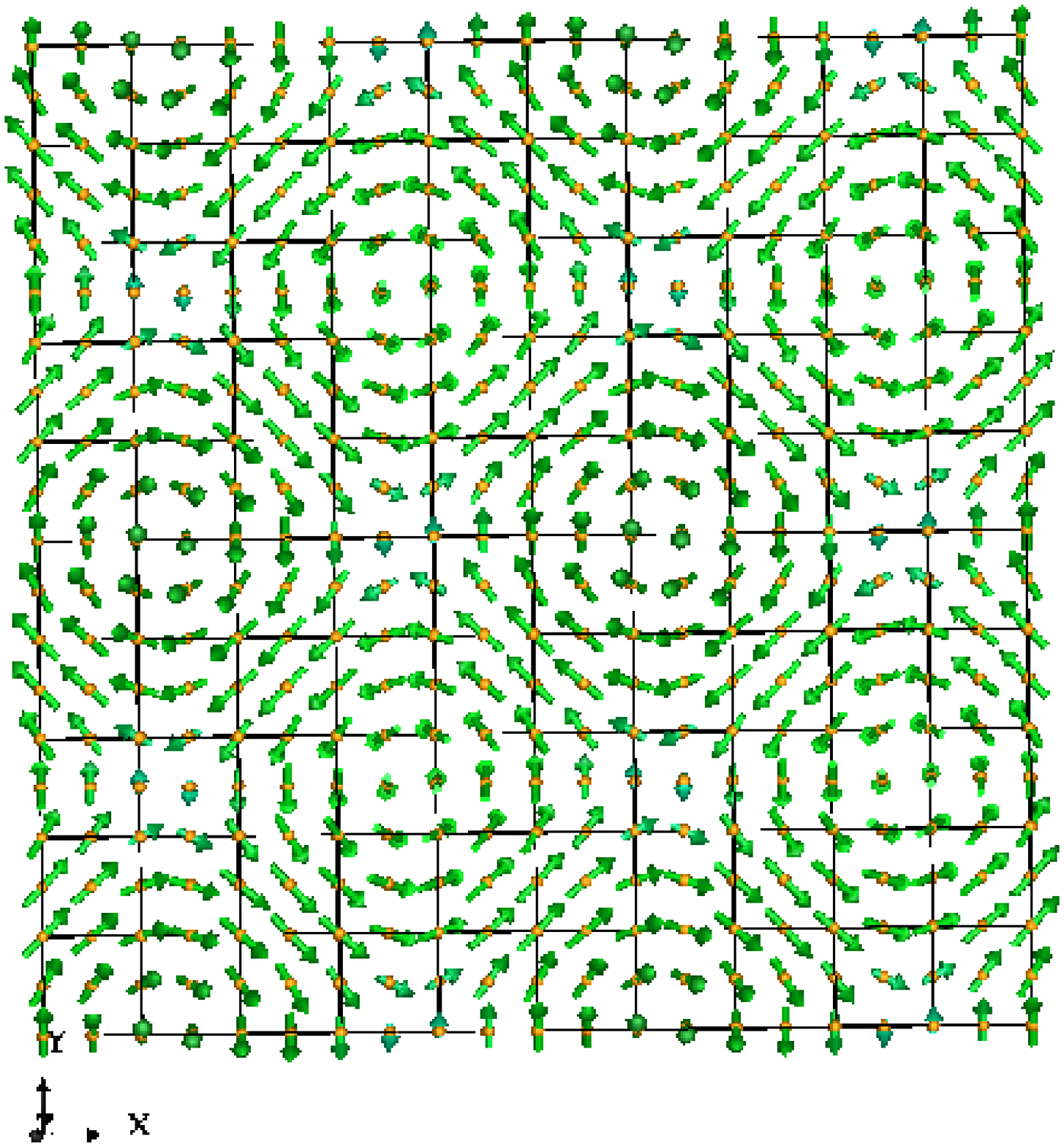} 
\vspace{-0.5cm} \\
(b)
\vspace{5mm} \\
\end{center}
\end{minipage}
\end{center}
\vspace{0cm}
\caption{Input structure of the 2$Q$ multiple helical spin density waves (2QH). \ \ (a) $q=1$, \ \ (b) $q=0.2$.
}
\label{fg2qhin}
\end{figure}
%
%

Note that the 2QH for $q=1$ shows a simple zigzag structure in which magnetic moments on the antiferromagnetic (AF) $x$-$y$ planes alternatively change their directions by $\pi/2$ with the translation $a/2$ along the $z$ axis (see Fig. \ref{fg2qhin}(a)).   The 2QH for small $q$, on the other hand, form the vortex-type particles with size $\lambda = a/q$ as shown in Fig. \ref{fg2qhin}(b).  The local magnetic moments (LM) at their margins are not antiparallel to those at the core position, but perpendicular to the LM at the core position.  Thus these are the half-skyrmions (merons) or the half-antiskyrmions (antimerons) depending on the sign of core polarization.  In fact, the topological number $Q_{N}$ of each vortex as a single skyrmion becomes $1/2$ or $-1/2$ according to the formula $Q_{N}=pq$, where the polarization number $p = \pm 1/2$ and the winding number $w = 1$ in the present case~\cite{merm79,naga13}.  The 2QH forms an  AF lattice consisting of the half-skyrmion and half-antiskyrmion (meron-antimeron) magnetic particles on the $x$-$y$ plane.  The 2QH in the long-wave regime ($i.e., q << 1$) is the same as the meron-antimeron structure found in the Co${}_{8}$Zn${}_{9}$Mn${}_{3}$ thin film system~\cite{yu18}.

\begin{figure}[htbp]
\begin{center}
\includegraphics[width=4cm]{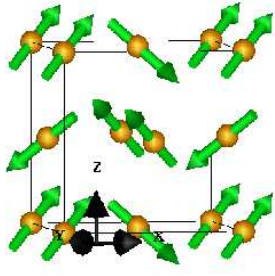} 
\end{center}
\vspace{0cm}
\caption{Input 3$Q$ multiple helical spin density waves (3QH) structure for $q=1$.
}
\label{fg3qhina}
\end{figure}

The 3QH with $q=1$ shows a noncollinear magnetic structure as shown in Fig. \ref{fg3qhina}.  The LM at the origin points to the $[111]$ direction.  The neighboring 3 LM on a $(111)$ plane at the face-centered positions form a small anticlockwise or clockwise vortex.  These
vortex planes stack alternatively along the $[111]$ direction.

\begin{figure}[htbp]
\begin{center}
\begin{minipage}{10.0cm}
\begin{center}
\includegraphics[width=10.0cm]{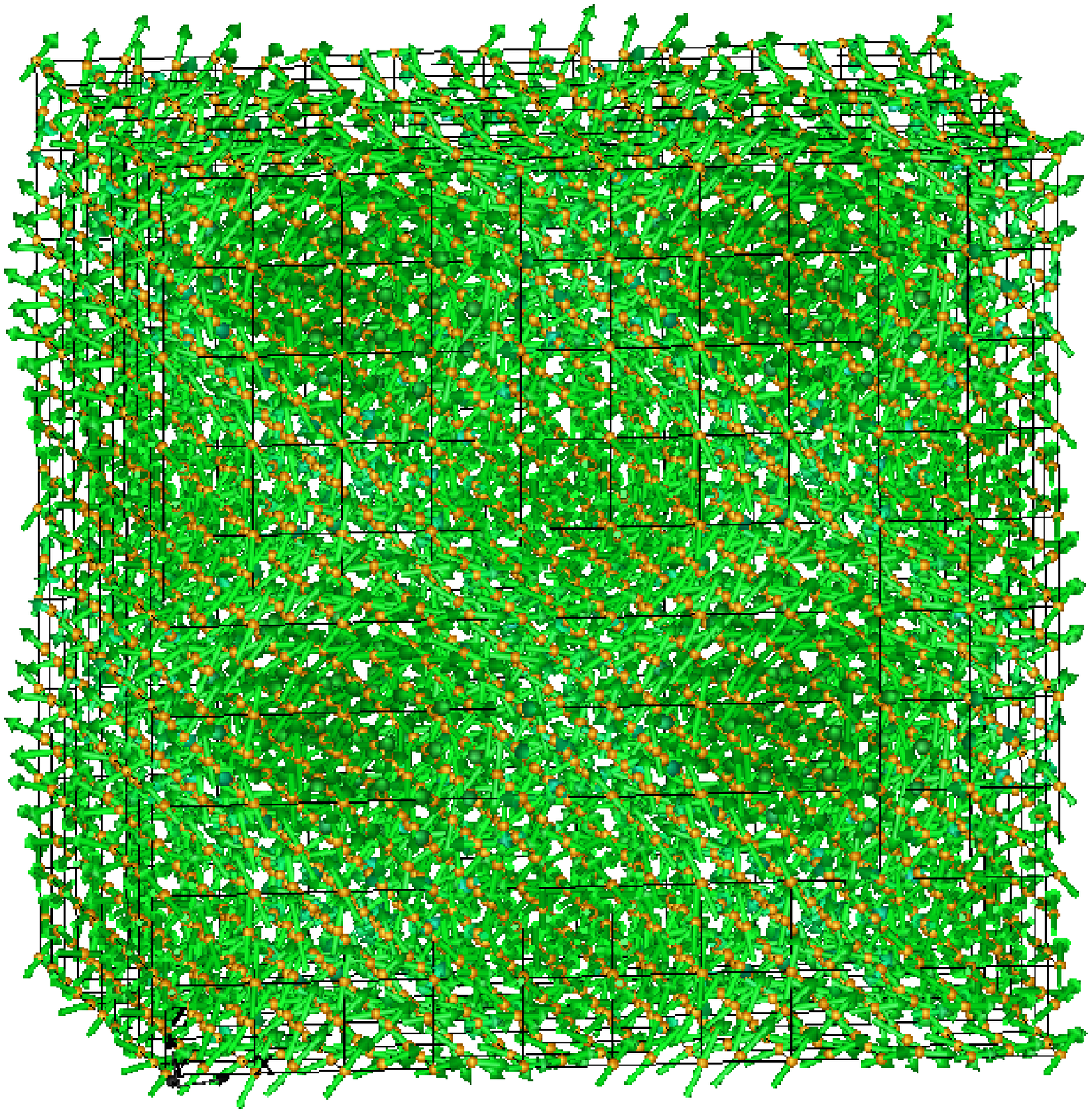} 
\vspace{-0.9cm} \\
(a)
\vspace{7mm} \\
\end{center}
\end{minipage}
\hspace{0.2cm}
\begin{minipage}{5cm}
\begin{center}
\vspace*{3.7cm}
\includegraphics[width=5.0cm]{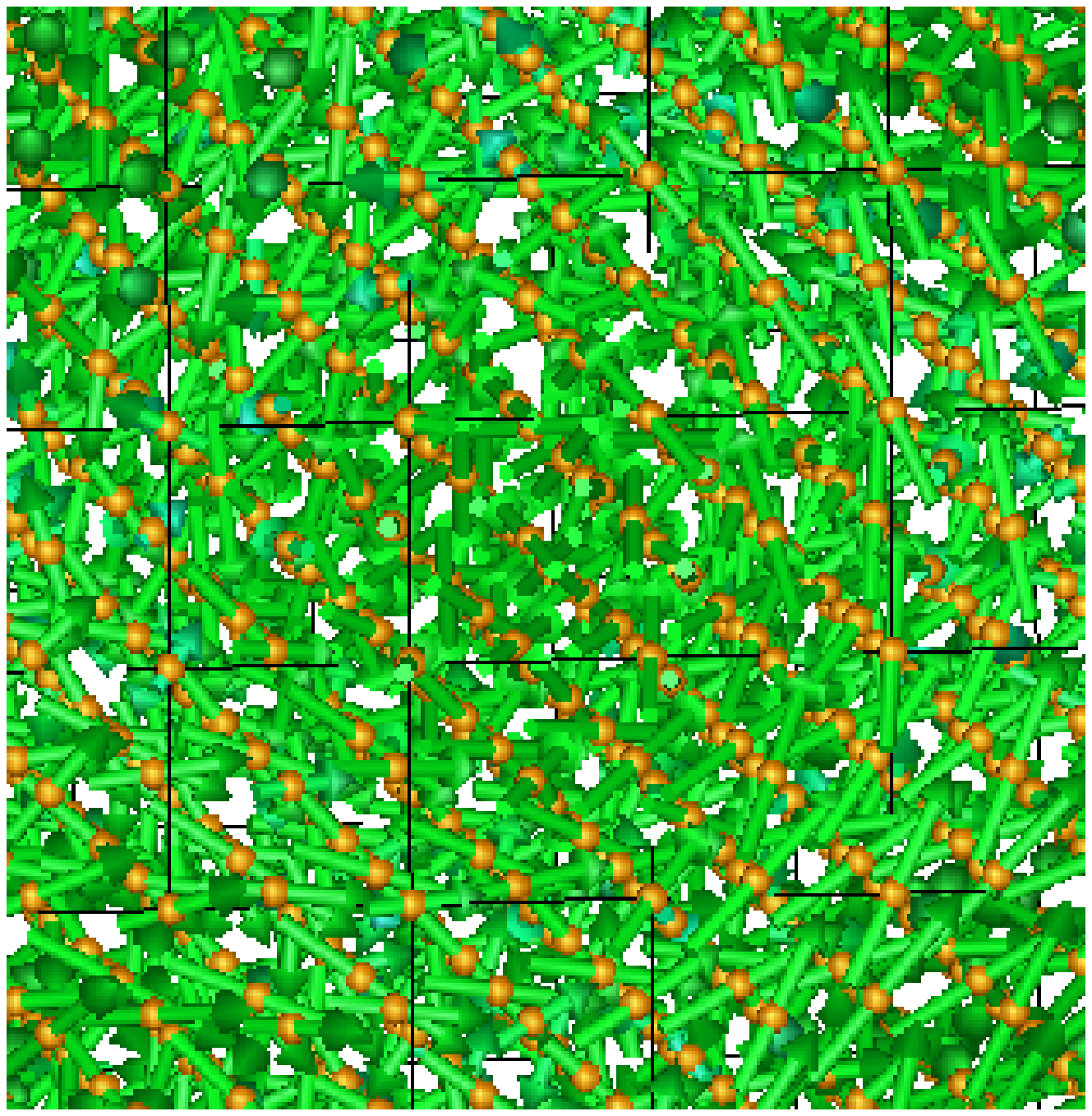} 
\vspace{-0.0cm} \\
(b) 
\vspace{5mm} \\
\end{center}
\end{minipage}
\end{center}
\vspace{0cm}
\caption{(a) Input 3QH structure for $q=0.2$.   \ \ (b) Enlarged view for showing a local spin structure for the central vortex.  The 3QH is regarded as a superposition of the 2QH and the 1QH.  On the $x$-$z$ plane, the 2QH vortex particles as shown in Fig. \ref{fg2qhin}(b) is influenced by the remaining 1QH ``westerlies'' $m\boldsymbol{k}$, so that the right-hand-side (left-hand-side) of the vortex flow is enhanced and the left-hand-side (right-hand-side) is weakened in the anticlockwise (clockwise) vortex.  These biased vortices are rotated with the translation along the $y$ axis   (see the text). Use a zoom-in tool to see more detailed structure.  
}
\label{fg3qhinb}
\end{figure}

When the wave number $q$ is small, the 3QH shows a vortex-type structure as shown in Fig. \ref{fg3qhinb}.  This structure can be understood as a superposition of the 2QH and the 1QH (3QH = 2QH + 1QH).  As has been mentioned, the 2QH forms the half-skyrmion and half-antiskyrmion vortex structure on the $x$-$y$ plane.  The remaining 1QH with $\bQ_{3}$ wave vector adds a uniform polarization 
$m \be_{\bot}$ to the 2QH on a $x$-$y$ plane, where 
$\be_{\bot} = \boldsymbol{i} \cos \bQ_{3} \! \cdot \! \boldsymbol{R}_{l} +
\boldsymbol{j} \sin \boldsymbol{Q}_{3} \! \cdot \! \boldsymbol{R}_{l}$.
Then the magnetic moments parallel to $\be_{\bot}$ in a 2QH vortex particle are enhanced, while those antiparallel to $\be_{\bot}$ are reduced, as in the typhoon with westerlies.  (This behavior is observed in the vortex particles on the $x$-$z$ plane in Fig. \ref{fg3qhinb}(b), where the ``westerlies'' is in the $z$ direction.)
The ``westerlies'' $m \be_{\bot}$ rotates along the $z$ axis with the period $\lambda = a/q$
.  Therefore the 3QH shows a vortex structure such that the 2QH vortex structure with the ``westerlies'' $m \be_{\bot}$ on the $x$-$y$ plane is twisted with the pitch $\lambda$ along the $z$ axis.  Needless to say, the same arguments hold true after making a cyclic exchange of $x$, $y$, and $z$, because Eq. (\ref{3qm}) for the 3QH structure is invariant for the cyclic transformation.  The input 3QH in the long-wave limit agrees with that used in the structure analysis of MnGe system~\cite{zhan16}.
%
%
%
%
%
%
%
%

When the wave vectors $q$ in the 2QH and 3QH are close to 1, these states show an AF-base vortex structures in which the ``amplitudes'' of local moments change their sign antiferromagnetically, as discussed in our last paper~\cite{kake18}.  The 3QH, for example, is expressed as follows:
\begin{align}
\langle \bm_{l} \rangle &= m_{\rm AF1}(\bR_{l}) \, \big(  \, 
\boldsymbol{j} \cos \tilde{\boldsymbol{q}}_{1} \! \cdot \! \boldsymbol{R}_{l} -
\boldsymbol{k} \sin \tilde{\boldsymbol{q}}_{1} \! \cdot \! \boldsymbol{R}_{l} \big)  \nonumber \\
&\ \ \ \ +
m_{\rm AF2}(\bR_{l}) \, \big(  \,
\boldsymbol{k} \cos \tilde{\boldsymbol{q}}_{2} \! \cdot \! \boldsymbol{R}_{l} -
\boldsymbol{i} \sin \tilde{\boldsymbol{q}}_{2} \! \cdot \! \boldsymbol{R}_{l} \big) \nonumber \\
&\ \ \ \ +
m_{\rm AF3}(\bR_{l}) \, \big(  \,
\boldsymbol{i} \cos \tilde{\boldsymbol{q}}_{3} \! \cdot \! \boldsymbol{R}_{l} -
\boldsymbol{j} \sin \tilde{\boldsymbol{q}}_{3} \! \cdot \! \boldsymbol{R}_{l} \big) \ ,
\label{af3qm}
\end{align}
where $m_{{\rm AF}n}(\bR_{l}) = m \cos (\hat{\boldsymbol{Q}}_{n} \! \cdot \! \boldsymbol{R}_{l})$, 
$\hat{\boldsymbol{Q}}_{1} = (1,0,0)$, $\hat{\boldsymbol{Q}}_{2} = (0,1,0)$, 
$\hat{\boldsymbol{Q}}_{3} = (0,0,1)$, and 
$\tilde{\boldsymbol{q}}_{n} = \hat{\boldsymbol{Q}}_{n} - \boldsymbol{Q}_{n}$.

It should be noted that the 2QH and the 3QH given by Eqs. (\ref{2qm}) $\sim$ (\ref{af3qm}) do not satisfy the GHF self-consistency.  We have to solve the GHF self-consistent  equations (\ref{chargeni}) and  (\ref{spinmi}) iteratively, starting from the structures given by Eqs. (\ref{2qm}) and (\ref{3qm}).

We also considered the paramagnetic state (P), the ferromagnetic state (F), the antiferromagnetic state (AF), the helical state (1QH), and conical state (C), in addition to the 2QH and 3QH structures.
Starting from these structures and varying the wave number $q$, we performed the self-consistent GHF calculations at zero temperature for a given $U$ and an electron number per site $n$,  and determined the stable magnetic structure comparing their total energies (see  Eq. (\ref{gener})).
It should be noted that the self-consistent output magnetic structures do not necessarily agree with the starting input magnetic structures, so that we obtain a variety of complex magnetic structures in the present approach.
%
%
\begin{figure}[htbp]
\begin{center}
\includegraphics[width=13cm]{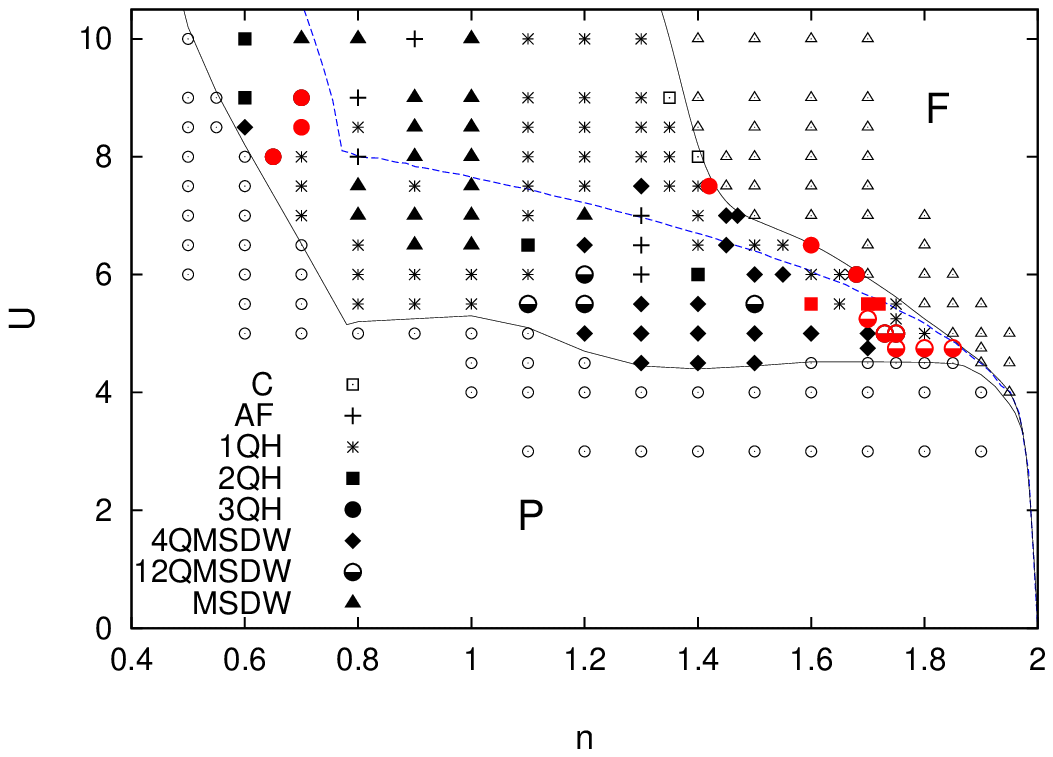}
\end{center}
\vspace{0cm}
\caption{Magnetic phase diagram on the $U$-$n$ plane showing various magnetic structures.  F: ferromagnetic state (open triangles), P: paramagnetic state (open circles), C: conical state, AF: anti-ferromagnetic state,  1QH: 1$Q$ helical state, 2QH: 2$Q$ multiple helical spin density wave  (2QMHSDW) state, 3QH: 3$Q$ MHSDW state, 4QMSDW: 4$Q$ multiple spin density wave (MSDW) state, 12QMSDW: 12$Q$ MSDW state, and MSDW: the other-type MSDW states.  The Coulomb interaction energy parameter $U$ is measured in unit of the nearest-neighbor transfer integral $|t|$.  The vortex type structures with long wave length are shown by red color.  The paramagnetic and ferromagnetic boundaries are shown by thin curves. The Stoner instability line is shown by the blue dashed line.  
}
\label{fgunpd}
\end{figure}
%
%

\subsection{Overview of the magnetic phase diagram}

Calculated magnetic phase diagram on the $U$-$n$ plane is presented in Fig. \ref{fgunpd}.  The Coulomb interaction energy parameter $U$ is measured in unit of $|t|$ here and hereafter. The ferromagnetic state (F) appears in the upper-right region of the phase diagram, while the paramagnetic state (P) is stabilized in the lower-left region, being consistent with the Stoner condition $U > 1/\rho(0)$, where $\rho(0)$ is the nonmagnetic DOS per atom per spin at the Fermi level.  The stability of F in the vicinity of $n=2$ and $U=0$ is originated in the divergence of the fcc DOS at the band edge.  Calculated F and P phase boundaries are consistent with those obtained by Igoshev {\it et al}~\cite{igos15}.  

Between the F and P boundaries, various magnetic structures appear.  These structures include the Conical structure (C), the Anti-Ferromagnetic structure (AF), the Helical structure (1QH), the 2$Q$ multiple Helical SDW structure (2QH), the 3$Q$ multiple Helical SDW (3QH), the 4$Q$ Multiple SDW (4QMSDW), the 12$Q$ Multiple SDW (12QMSDW), and the other Multiple SDW (MSDW) states.

Note that we have classified the region between the F and P states according to the multiplicity of the SDW's since main purpose of the present work is to search the skyrmion-type vortex structures in the itinerant electron system from a viewpoint of the MSDW.  In each class of structures, different phases may appear, but we did not determine in details their phase boundaries in the present work.

We obtained the conical state (C) in the vicinity of the ferromagnetic boundary.  Their wave numbers are given by $q=0.4$ for $U=8$ and $q=0.3$ for $U=9$, respectively.  The AF state of the first kind (AFI) with the wave vector $\bQ = (0,0,1)$ is stabilized at $(n,U) = (0.8, 8)$, $(0.8, 9)$, and $(0.9, 10)$, while the AF state of the second kind (AFII) with  $\bQ = (0.5, 0.5, 0.5)$ is stabilized at $(1.3, 6)$, $(1.3, 6.5)$, and $(1.3, 7)$.
%
%
\begin{figure}[htbp]
\begin{center}
\includegraphics[width=12cm]{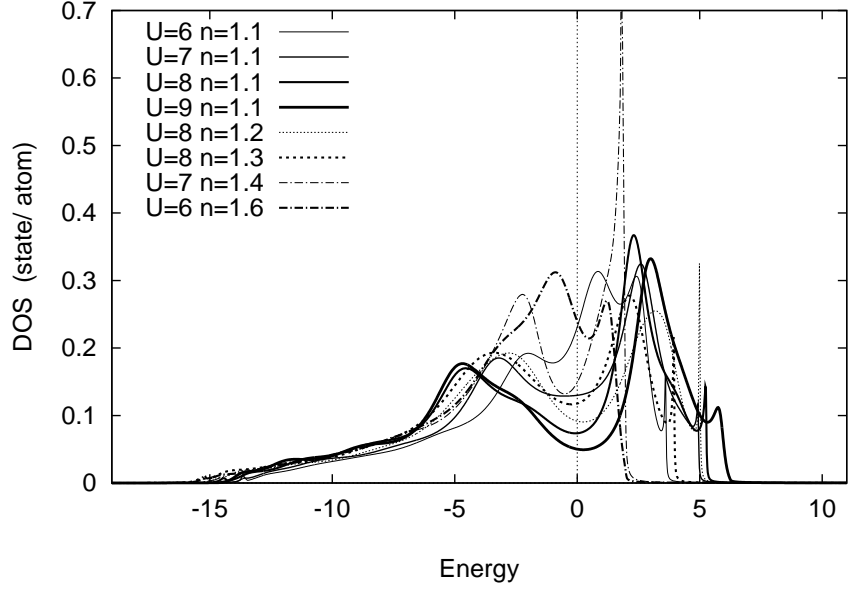}
\end{center}
\vspace{0cm}
\caption{Densities of States (DOS) for the 1$Q$ helical structure (1QH) in various $(U,n)$ points.  Energy is measured from the Fermi level.
}
\label{fg1qhds}
\end{figure}
%
%

The 1QH state with $\bQ = (0,0,q)$ appears in a wide range of region between the F and P boundaries.  The wave number $q$ in general shows the maximum around $n=0.8$ and monotonically decreases with increasing $n$ when $U$ is fixed.  For example, we obtained $q=1.0$ at $n=0.8$, $q=0.8$ at $n=1.1$, $q=0.5$ at $n=1.2$ and $1.3$ in the case of $U=8$.  We note that the 1QH at $(1.4, 6.5)$, $(1.4, 7)$, $(1.35, 7.5)$, and $(1.35, 8)$ have the wave vector $\bQ = (0.4, 0.4, 0.4)$, being consistent with the neighboring AFII with $\bQ = (0.5,0.5,0.5)$ at $(1.3, 7)$.
Usually, the Fermi level of the 1QH is located near the dip or valley of the DOS, as shown in Fig. \ref{fg1qhds}, indicating the stability due to the kinetic energy gain.

The other magnetic structures we found are the multiple SDW.  In particular, we found the vortex-type magnetic structures with small $q$, which are indicated by the red color in Fig. \ref{fgunpd}.  These structures are found around the Stoner instability line drawn by the dashed blue curve.  We present the MSDW in more details in the following subsections.

\subsection{2$Q$ multiple helical SDW}

\begin{figure}[htbp]
\begin{center}
\includegraphics[width=10cm]{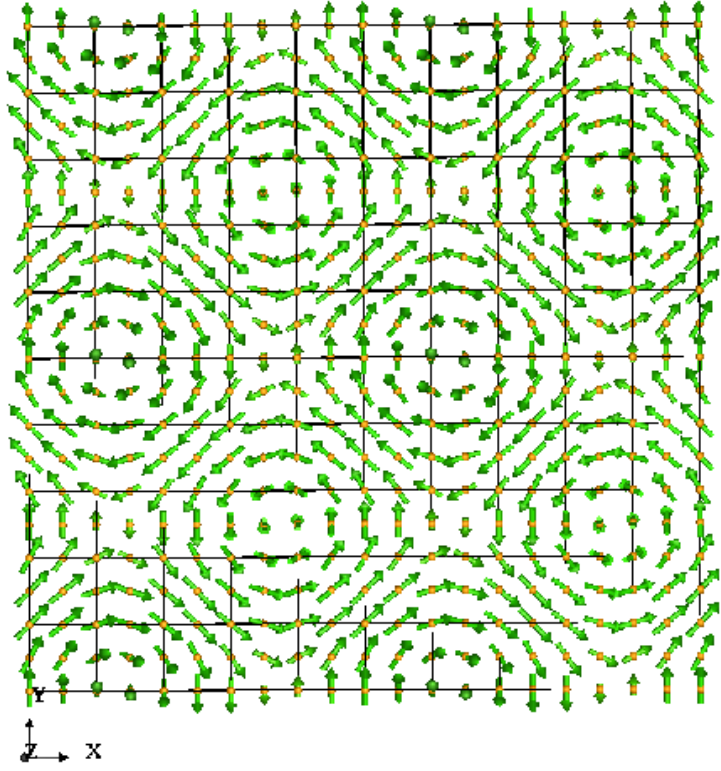}
\end{center}
\vspace{0cm}
\caption{Self-consistent 2QH ($q=0.2$) structure obtained at $(n,U)=(1.7,5.5)$.
}
\label{fg2qh}
\end{figure}

The self-consistent 2$Q$ multiple helical SDW (2QH) with small wave number $q=0.2$ has been found in a narrow region $n = 1.70 \sim 1.73$ and $U = 5.5$ as shown in Fig. \ref{fg2qh}.  The magnetic structure is close to the input ideal 2QH with vortex structures.  We can observe the AF arrangement of the half-skyrmion and half-antiskyrmion particles with size $\lambda = a/q = 5a$ on the $x$-$y$ plane.
%
%
%
%

\begin{figure}[htbp]
\begin{center}
\begin{minipage}{7cm}
\begin{center}
\vspace*{-1.0cm}
\includegraphics[width=7cm]{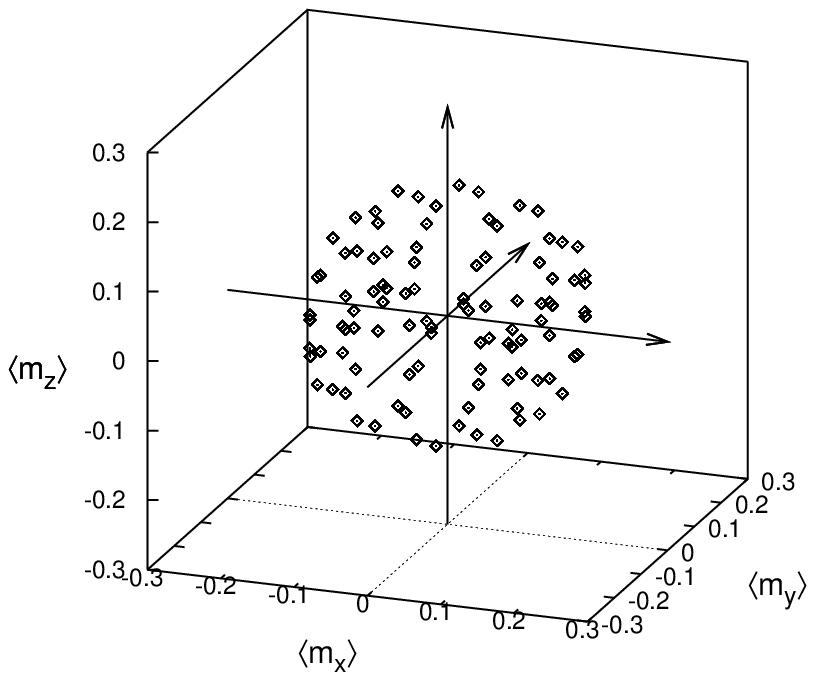} 
\vspace{-1.5cm} \\
(a) 
\vspace{5mm} \\
\end{center}
\end{minipage}
\hspace{0.7cm}
\begin{minipage}{7cm}
\begin{center}
\vspace*{0cm}
\includegraphics[width=7cm]{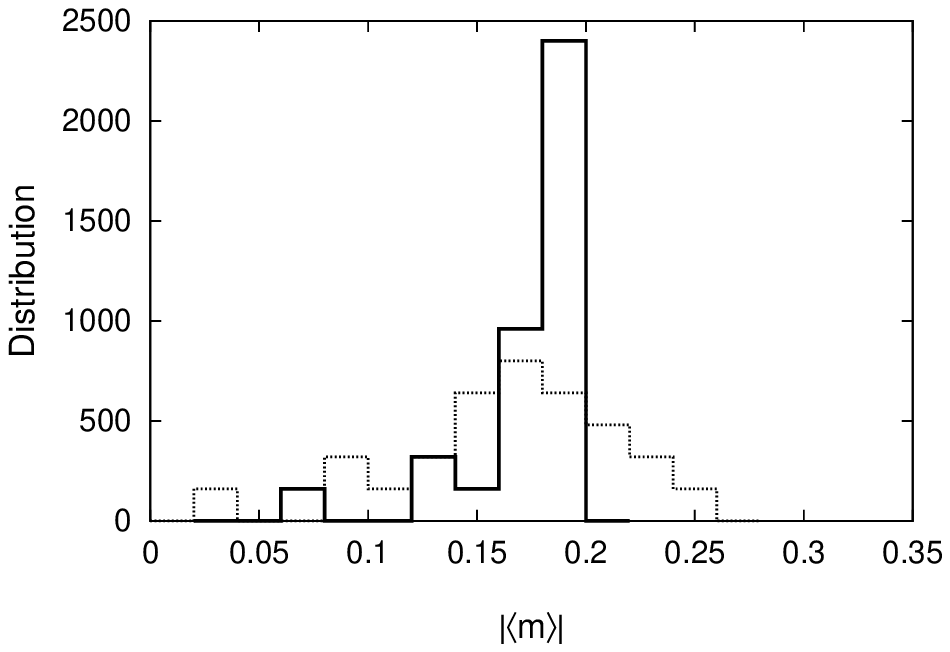} 
\vspace{-0.5cm} \\
(b)
\vspace{5mm} \\
\end{center}
\end{minipage}
\end{center}
\vspace{0cm}
\caption{(a) Local moment (LM) distribution for self-consistent 2QH ($q=0.2$) at $(n,U)=(1.7,5.5)$. 4000 LM in the $10 \times 10 \times 10$ cluster are plotted in the same $(\langle m_{x} \rangle, \langle m_{y} \rangle, \langle m_{z} \rangle)$ space. \ \  (b) Amplitude distribution of LM for the same 2QH.  The dotted line expresses the distribution for the ideal 2QH.
}
\label{fg2qhlm}
\end{figure}

We made the Fourier analysis of the magnetic structure using the relation $\langle \bm_{l} \rangle = \sum_{\bq} \bm(\bq) \exp (i\bq \cdot \bR_{l})$. The principal terms of the self-consistent 2QH are two elliptical-helical waves with amplitudes $|\bm(\bQ_{1})| = |\bm(\bQ_{2})| = 0.087$.
\begin{align}
\langle \bm_{l} \rangle &= m_{1} 
\boldsymbol{j} \cos \boldsymbol{Q}_{1} \! \cdot \! \boldsymbol{R}_{l} +
m_{2} \boldsymbol{k} \sin \boldsymbol{Q}_{1} \! \cdot \! \boldsymbol{R}_{l}  \nonumber \\
& \ +
m_{2} \boldsymbol{k} \cos \boldsymbol{Q}_{2} \! \cdot \! \boldsymbol{R}_{l} +
m_{1} \boldsymbol{i} \sin \boldsymbol{Q}_{2} \! \cdot \! \boldsymbol{R}_{l}  \ .
\label{2qeh}
\end{align}
Here $m_{1}=0.075$, $m_{2}=0.044$, $\bQ_{1}=(q,0,0)$, $\bQ_{2}=(0,q,0)$, and $q=0.2$.
They are accompanied by additional 4$Q$-MSDW with smaller amplitudes 
$|\bm(\bQ)| = 0.004$ and the wave vectors $\bQ = (\pm 0.2, \pm 0.4, 0)$ and 
$(\pm 0.4, \pm 0.2, 0)$.
Because of the deviation from the ideal 2QH, the LM distribution becomes more spherical  and the amplitude modulation is suppressed as compared with the ideal 2QH as shown in Fig. \ref{fg2qhlm}.
Note that the suppression of the amplitude modulation reduces the Coulomb energy loss, and thus is favorable for the stability of the self-consistent 2QH structure.
%
%
%
%

\begin{figure}[htbp]
\begin{center}
\includegraphics[width=10cm]{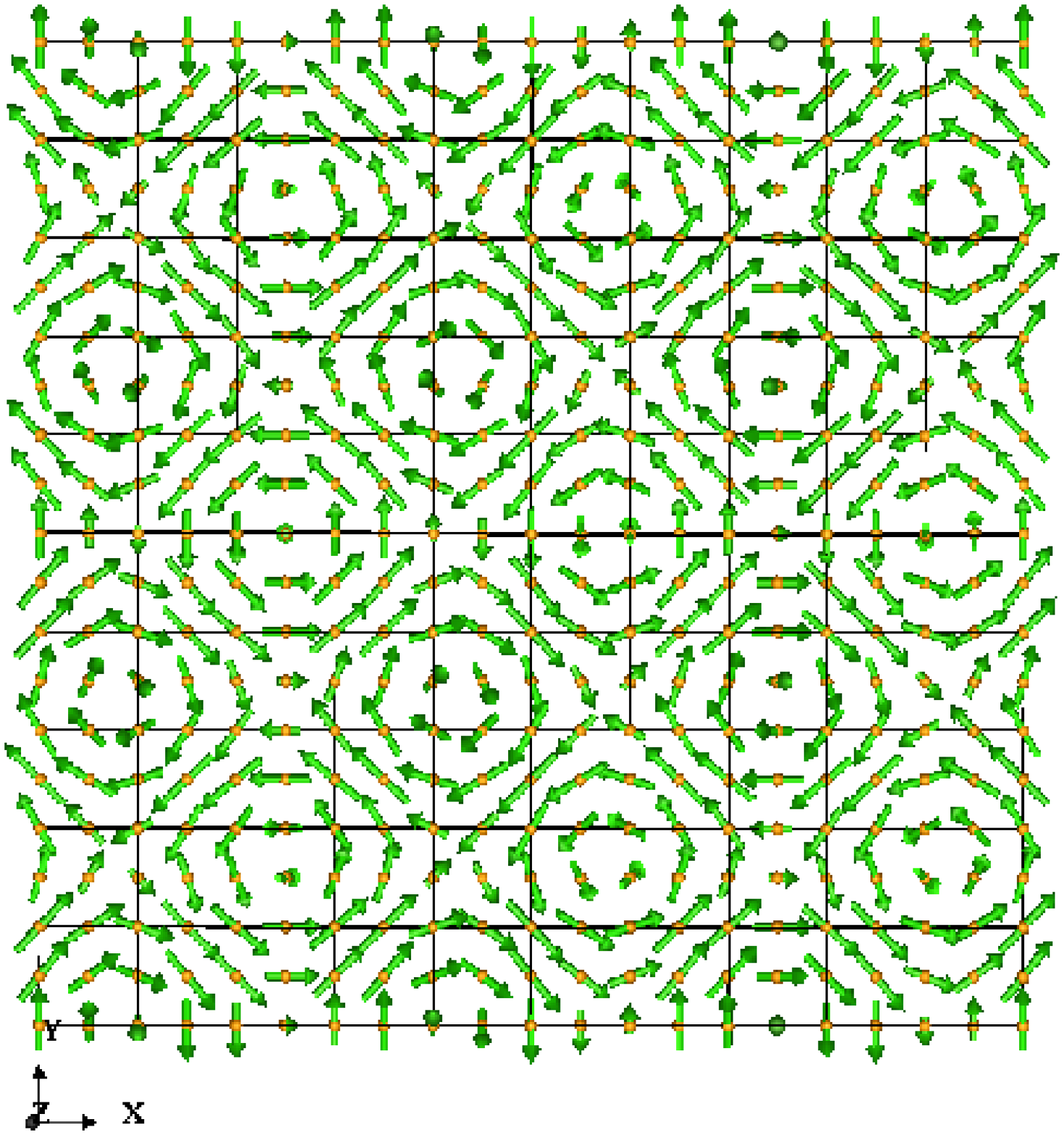}
\end{center}
\vspace{0cm}
\caption{Self-consistent 2QH ($q=0.3$) structure obtained at $(n,U)=(1.6,5.5)$.
}
\label{fg2qh03}
\end{figure}

\begin{figure}[htbp]
\begin{center}
\includegraphics[width=9cm]{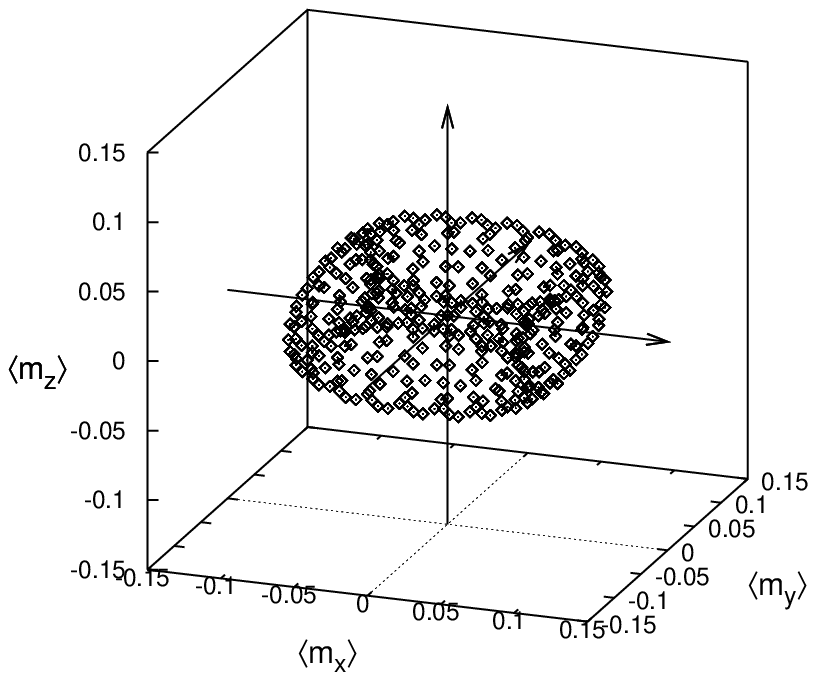}
\end{center}
\vspace{0cm}
\caption{Local moment distribution for the self-consistent 2QH ($q=0.3$) at $(n,U)=(1.6,5.5)$.
}
\label{fg2qhlm03}
\end{figure}

The same type of 2QH with larger $q \, (=0.3)$ appears at $(1.6, 5.5)$.  As shown in Fig. \ref{fg2qh03},  the half-skyrmion particles are not exactly equivalent to each other in this case, because the particle size $\lambda = 10a/3$ is not commensurate with the lattice.  The amplitudes $m_{1}$ and $m_{2}$ in Eq. (\ref{2qeh}) are smaller and more anisotropic;  $m_{1}=0.040$ and $m_{2}=0.015$.  Thus, we obtain a pillow-type LM distribution (see Fig. \ref{fg2qhlm03}).
%
%
%
%
%
%
%
%

\begin{figure}[htbp]
\begin{center}
\includegraphics[width=10cm]{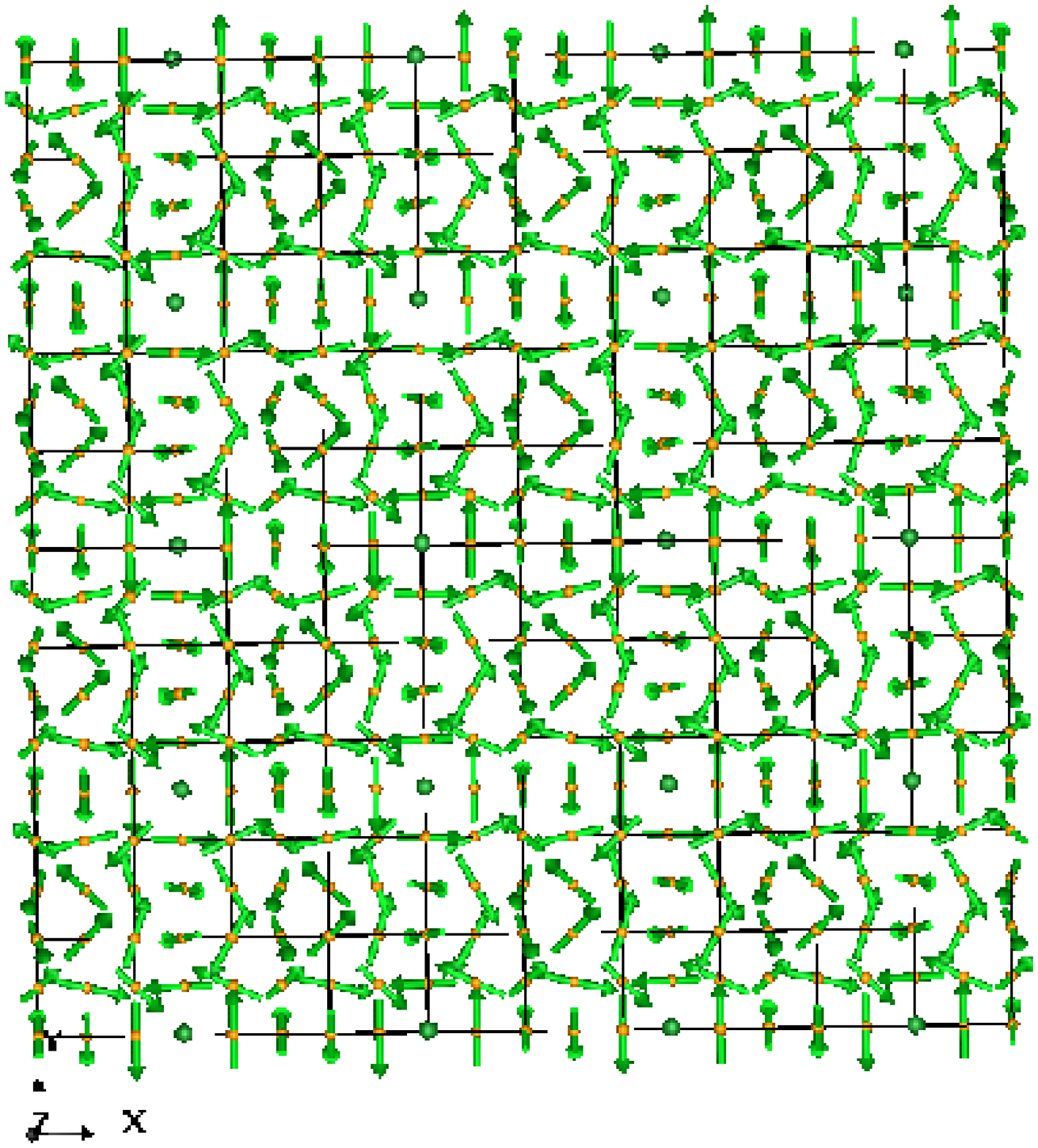}
\end{center}
\vspace{0cm}
\caption{Self-consistent 2QH ($q=0.4$) structure obtained at $(n,U)=(1.4,6)$.
}
\label{fg2qh04}
\end{figure}

When the electron number $n$ is decreased  further, we find a 2QH state with larger wave number $q=0.4$ at (1.4, 6).  According to the Fourier analysis the main terms are the 2$Q$ elliptical-helical SDW with the wave vectors $\bQ_{1} = (q, q, 0)$ and  $\bQ_{2} = (q, -q, 0)$.
\begin{align}
\langle \bm_{l} \rangle &= m_{1} \be_{1} \cos \boldsymbol{Q}_{1} \! \cdot \! \boldsymbol{R}_{l} +
m_{2} \be_{2} \sin \boldsymbol{Q}_{1} \! \cdot \! \boldsymbol{R}_{l}  \nonumber \\
&\ \ \ \ +
m_{1} \be_{3} \cos \boldsymbol{Q}_{2} \! \cdot \! \boldsymbol{R}_{l} +
m_{2} \be_{4} \sin \boldsymbol{Q}_{2} \! \cdot \! \boldsymbol{R}_{l}  \ .
\label{2qeh2}
\end{align}
Here $m_{1}=0.138$, $m_{2}=0.105$, $\be_{1} = (-0.68, 0.42, -0.60)$, $\be_{2} = (0.23, -0.79, -0.56)$, $\be_{3} = (0.68, 0.42, -0.60)$, and $\be_{4} = (-0.23, -0.79, -0.56)$.  The polarization unit vectors $\be_{1}$ and $\be_{2}$ ($\be_{3}$ and $\be_{4}$) are approximately orthogonal each other.  The 4$Q$-MSDW with $\bQ = (0.4, \pm 0.8, 0)$ and 
$(0.8, \pm 0.4, 0)$ and the 2$Q$-MSDW with $\bQ = (0.4, 0, 0)$ and $(0.0, 0.4, 0)$ appear as the next and third satellite contributions in the Brillouin zone of the wave vector.  Because $q \sim 1$ ({\it i.e.}, size $\lambda \sim a$) and the satellite 4$Q$- and 2$Q$-MSDW contributions are not negligible, the half-skyrmion picture no longer holds true in this case.  Instead we observe small and broken vortices with size $\lambda \sim a$ as seen in Fig. \ref{fg2qh04}.

The 2QH at $(n,U) = (1.1, 6.5)$ has the same form as Eq. (\ref{2qeh}) at $(1.7, 5.5)$, but the wave number is $q=0.6$, which is larger as compared with $q=0.2$ at $(1.7, 5.5)$, so that it consists of tiny and broken vortices with $\lambda=10a/6$.
%
%
%
%

The other 2QH states appear at $(0.6, 9)$ and $(0.6, 10)$ on the $U$-$n$ plane.  These structures do not show any vortex-type structure.  They are expressed as
\begin{align}
\langle \bm_{l} \rangle &= -m (
\boldsymbol{j} \cos \boldsymbol{Q}_{1} \! \cdot \! \boldsymbol{R}_{l} +
\boldsymbol{k} \sin \boldsymbol{Q}_{1} \! \cdot \! \boldsymbol{R}_{l} ) \nonumber \\
&\ \ \ \ -
m ( \boldsymbol{j} \cos \boldsymbol{Q}_{2} \! \cdot \! \boldsymbol{R}_{l} -
\boldsymbol{k} \sin \boldsymbol{Q}_{2} \! \cdot \! \boldsymbol{R}_{l} )  \ ,
\label{2qh3}
\end{align}
with $\bQ_{1} = (0.5, 1, 0)$ and $\bQ_{2} = (0.5, 0, 1)$.
There, the AF structure on the $y$-$z$ plane rotates by $\pi/2$ with a translation $a/2$ along the $x$ axis.
%
%
\begin{figure}[htbp]
\begin{center}
\includegraphics[width=12cm]{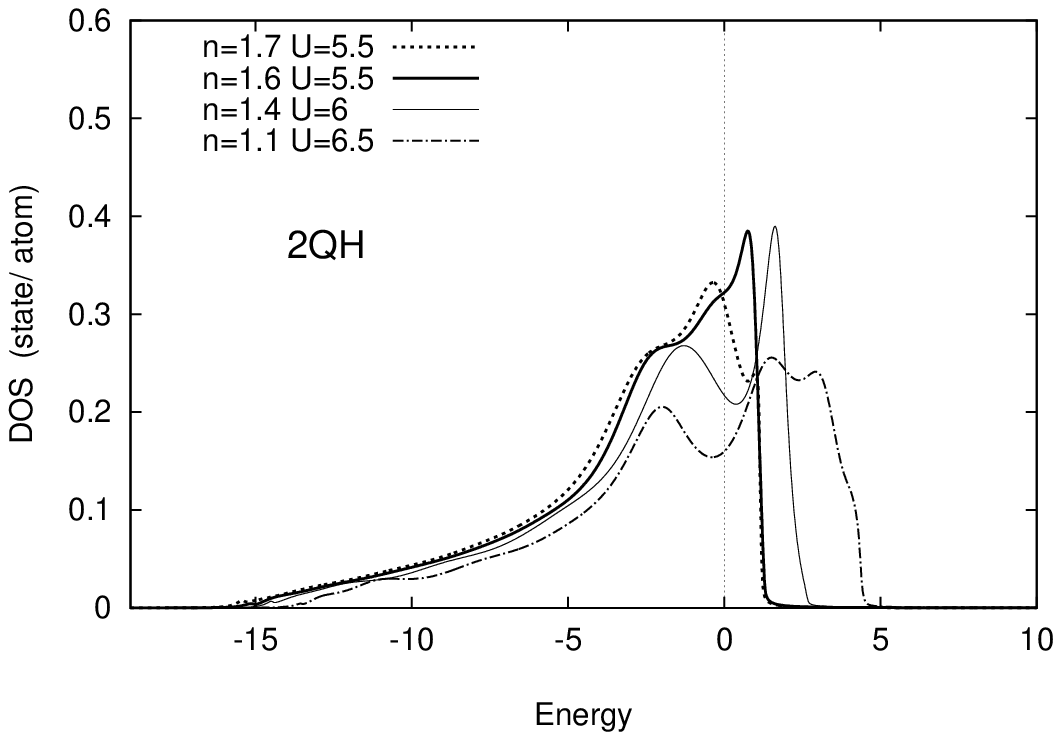}
\end{center}
\vspace{0cm}
\caption{DOS for various self-consistent 2QH.
}
\label{fg2qhds}
\end{figure}
%
%

We have examined the DOS of the vortex-type 2QH to understand the origin of their stability.  These DOS have a dip near the Fermi level as shown in Fig. \ref{fg2qhds}.  The 2QH for $(n,U) = (1.4, 6)$ and $(1.1, 6.5)$ are stabilized by a clear dip at the Fermi level.
The dip contributions to the stability are subtle for the remaining 2QH.  
Comparing the total energy between the 2QH and the 1QH with the same $q$, we examined the stability of the 2QH with vortex structure.  The total energy consists of the kinetic energy (the 2nd term at the rhs of Eq. (\ref{gener})), the charge potential term (the first and third terms), and the exchange energy term (the last term at the rhs of Eq. (\ref{gener})).
Hereafter we adopt the energy per atom in unit of $|t|$.  In the case of the vortex-type 2QH $(q=0.2)$ at $(1.7, 5.5)$,  the exchange energy gain $0.008$ due to the reduction of spin polarization approximately cancels with the charge-potential energy loss $0.007$, and the kinetic energy gain 0.007 due to the dip formation stabilizes the 2QH, when the 2QH is compared with 1QH.  For the 2QH $(q=0.3)$ at $(1.6, 5.5)$, we find that the exchange energy loss $0.006$ due to the change of the amplitudes of LM cancels with the charge-potential energy gain $0.004$, and the kinetic energy gain $0.006$ again contributes to the stability of the 2QH.

\subsection{3$Q$ multiple helical SDW}

\begin{figure}[htbp]
\begin{center}
\begin{minipage}{10.0cm}
\begin{center}
\includegraphics[width=10.0cm]{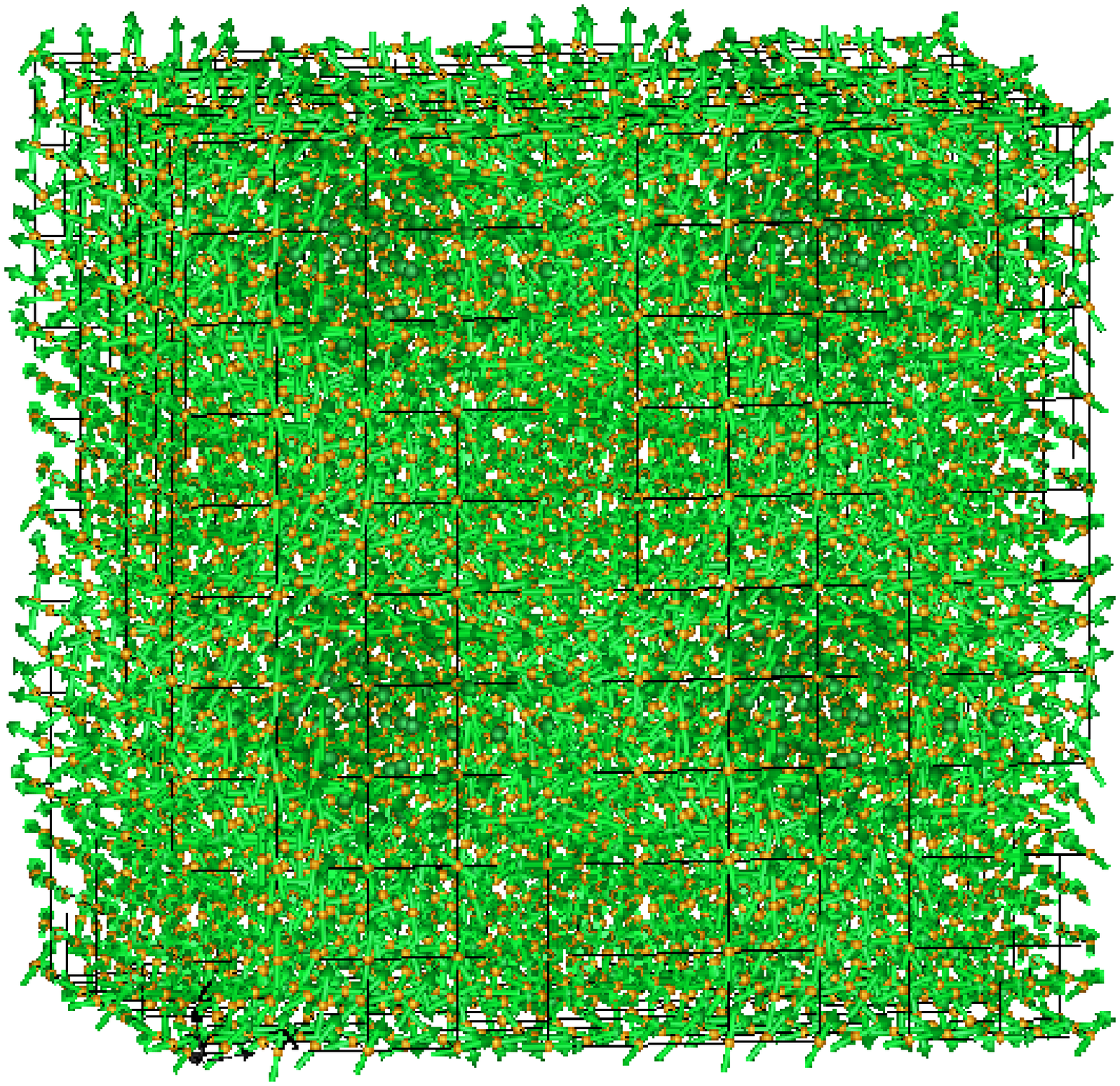} 
\vspace{-0.5cm} \\
(a)
\vspace{5mm} \\
\end{center}
\end{minipage}
\hspace{0.2cm}
\begin{minipage}{5cm}
\begin{center}
\vspace*{4.1cm}
\includegraphics[width=5cm]{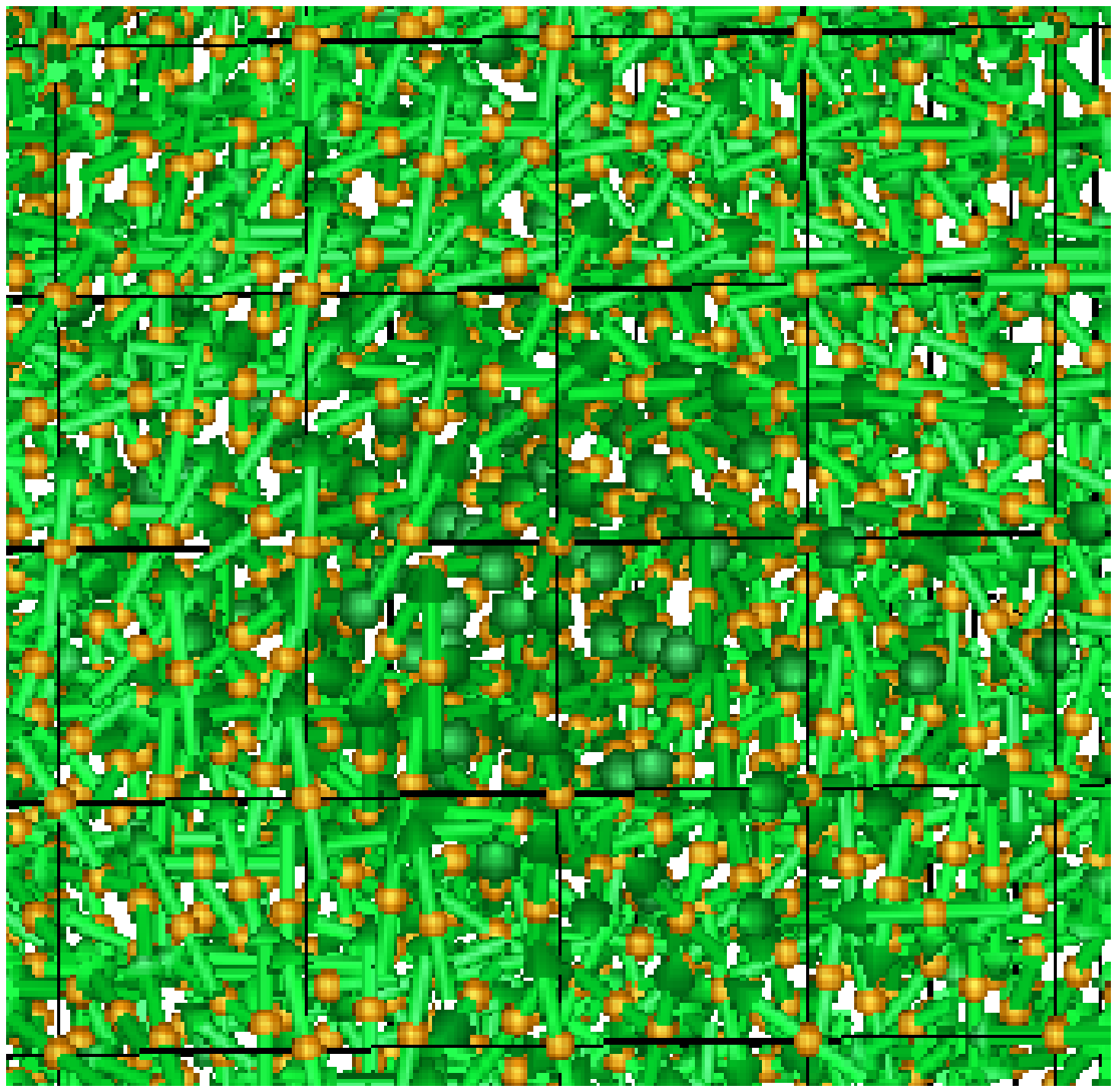} 
\vspace{0.0cm} \\
(b) 
\vspace{5mm} \\
\end{center}
\end{minipage}
\end{center}
\vspace{0cm}
\caption{(a) Self-consistent 3QH ($q=0.2$) structure at $(n,U)=(1.68,6)$. \ \ (b) Enlarged view for showing a local spin structure.  A clockwise vortex structure is observed.  Note that the left-hand-side is enhanced by the ``westerlies'' effects.  Use a zoom-in tool to see more detailed structure.
}
\label{fg3qh}
\end{figure}

\begin{figure}[htbp]
\begin{center}
\includegraphics[width=10cm]{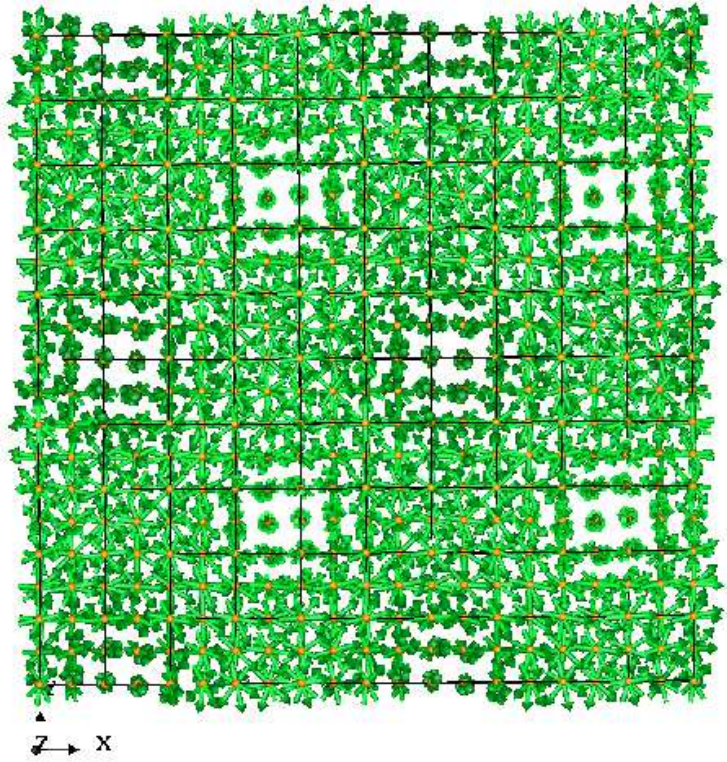}
\end{center}
\vspace{0cm}
\caption{Self-consistent 3QH ($q=0.2$) projected onto $x$-$y$ plane at $(n,U)=(1.68,6)$.  On the $x$-$y$ plane, the circular vortex shape of the ideal 3QH is changed to a square-like vortex shape due to additional 4$Q$-MSDW satellite contributuions.
}
\label{fg3qhxy}
\end{figure}

\begin{figure}[htbp]
\begin{center}
\begin{minipage}{7cm}
\begin{center}
\vspace*{-1.0cm}
\includegraphics[width=7cm]{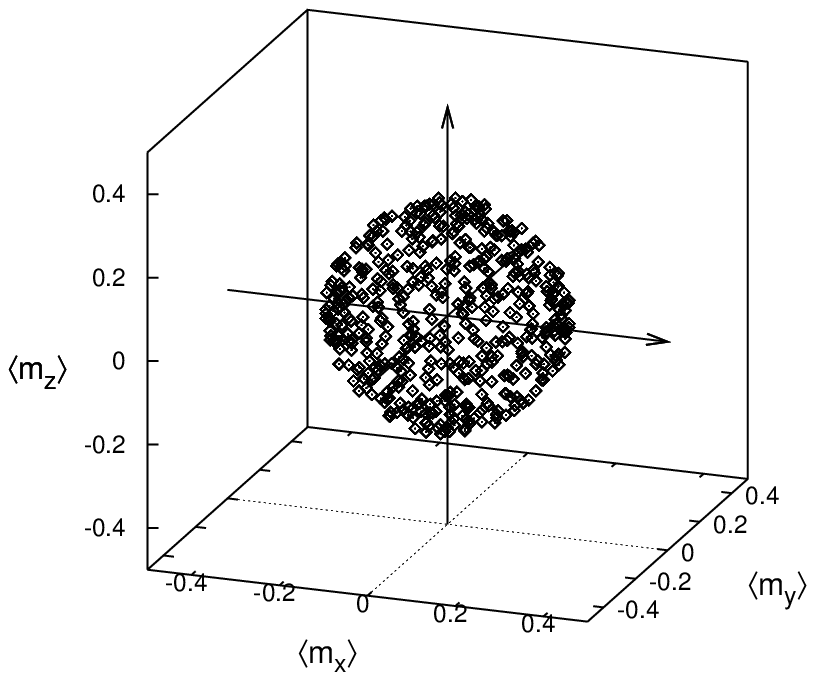} 
\vspace{-1.5cm} \\
(a) 
\vspace{5mm} \\
\end{center}
\end{minipage}
\hspace{0.7cm}
\begin{minipage}{7cm}
\begin{center}
\vspace*{0cm}
\includegraphics[width=7cm]{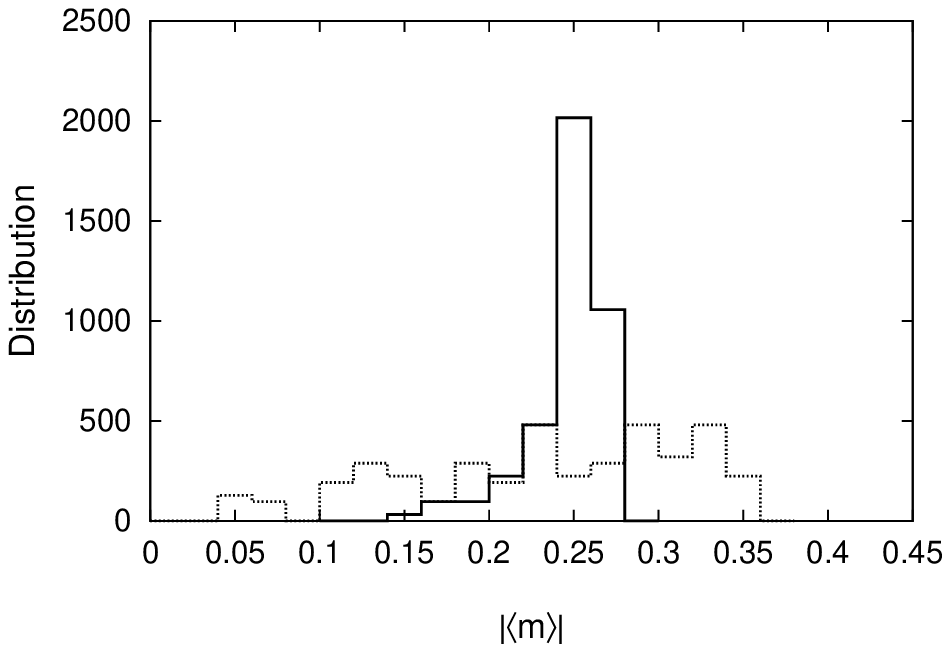} 
\vspace{-0.5cm} \\
(b)
\vspace{5mm} \\
\end{center}
\end{minipage}
\end{center}
\vspace{0cm}
\caption{(a) Local moment distribution for the self-consistent 3QH ($q=0.2$) at $(n,U)=(1.68,6)$ \ \  (b) Amplitude distribution of LM for the same 3QH.  Dotted histogram is the distribution for the ideal 3QH.
}
\label{fg3qhlm}
\end{figure}

The triple-$Q$ multiple helical SDW (3QH) with the long wave length have been found at $(n, U)=(1.68, 6)$, $(1.6, 6.5)$, and $(1.42, 7.5)$.  As shown in Fig. \ref{fg3qh},  the 3QH at $(1.68, 6)$ shows the vortex-type structure with particles size $\lambda=5a$ ($i.e.$, $q=0.2$) on the $x$-$z$ plane.  Each vortex particle is surrounded by the antivortex particles with opposite spin polarization, and is twisted along the $y$ axis as discussed in Sect. 3.1.  The Fourier analysis verifies that the vortex-type structure is formed by the 3QH with $|\bm(\bQ)| = 0.095$ and $q=0.2$.  But, in addition to the 3QH main contribution, there are satellite contributions of the 4$Q$-MSDW with the amplitude $|\bm(\bQ)| = 0.028$ and the wave vectors $\bQ = (\pm 0.2, \pm 0.2, \pm 0.2)$.  Due to these additional 4$Q$-MSDW, the circular vortex shape in the ideal 3QH (see Fig. \ref{fg3qhinb}(b)) changes to a square-like vortex shape on the $x$-$y$ plane as shown in Fig. \ref{fg3qhxy}.  Moreover, they suppress the amplitude fluctuations and produce a spherical LM distribution as shown in Fig. \ref{fg3qhlm}.

The 3QH at $(1.6, 6.5)$ and $(1.42, 7.5)$ show the same vortex type structure, whose principal terms are expressed by Eq. (\ref{3qm}) with $q=0.3$, and $m=0.083$ (for the former) and $0.134$ (for the latter).  The structure is again accompanied by the satellite 4$Q$-MSDW with $|\bm(\bQ)| = 0.033$ and the wave vectors $\bQ = (\pm 0.3, \pm 0.3, \pm 0.3)$. 
%
%
%
%
%
%
%
%
%
%
%
%

\begin{figure}[htbp]
\begin{center}
\includegraphics[width=12cm]{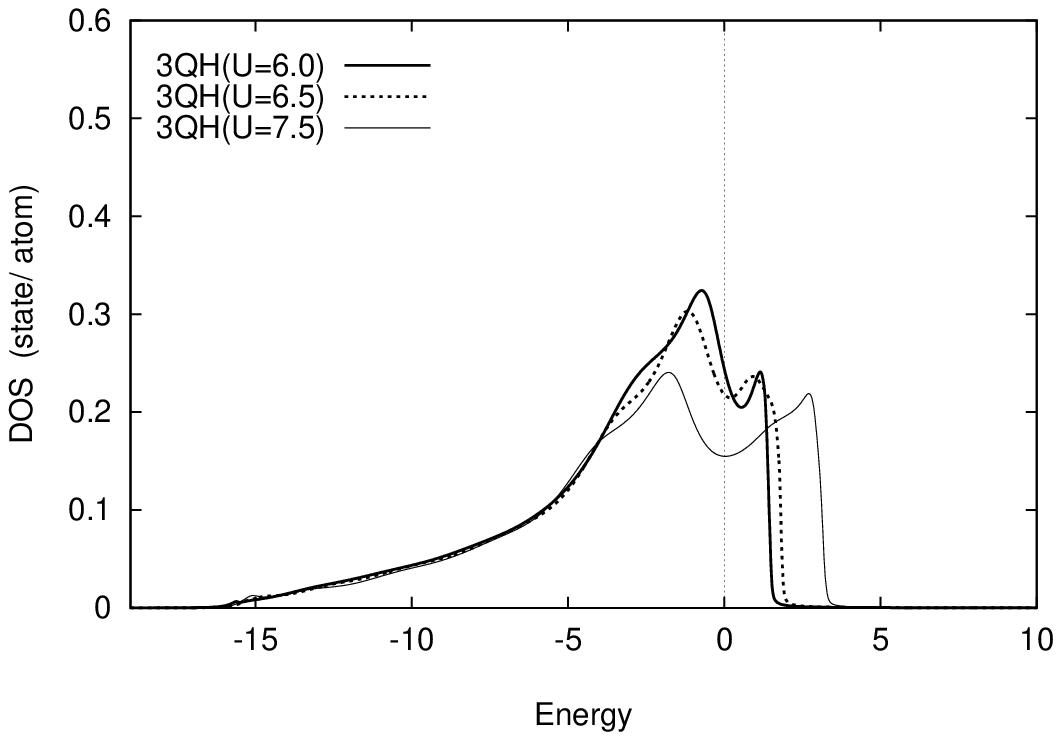}
\end{center}
\vspace{0cm}
\caption{Self-consistent DOS for 3QH with $q=0.2$ at $(n,U)=(1.68,6)$, 3QH with $q=0.3$ at $(1.6, 6.5)$, and 3QH with $q=0.3$ at $(1.42, 7.5)$.
}
\label{fg3qhds}
\end{figure}

The DOS for the 3QH show a sharp dip near the Fermi level as shown in Fig. \ref{fg3qhds}.  Associated kinetic energy gain is the origin of their stability.  In the case of $(n, U) = (1.68, 6)$, for example, the kinetic energy gain yields the total energy gain $0.019$ when the 3QH is compared with the 1QH.
%
%
%
%

\begin{figure}[htbp]
\begin{center}
\begin{minipage}{9.5cm}
\begin{center}
\includegraphics[width=9.5cm]{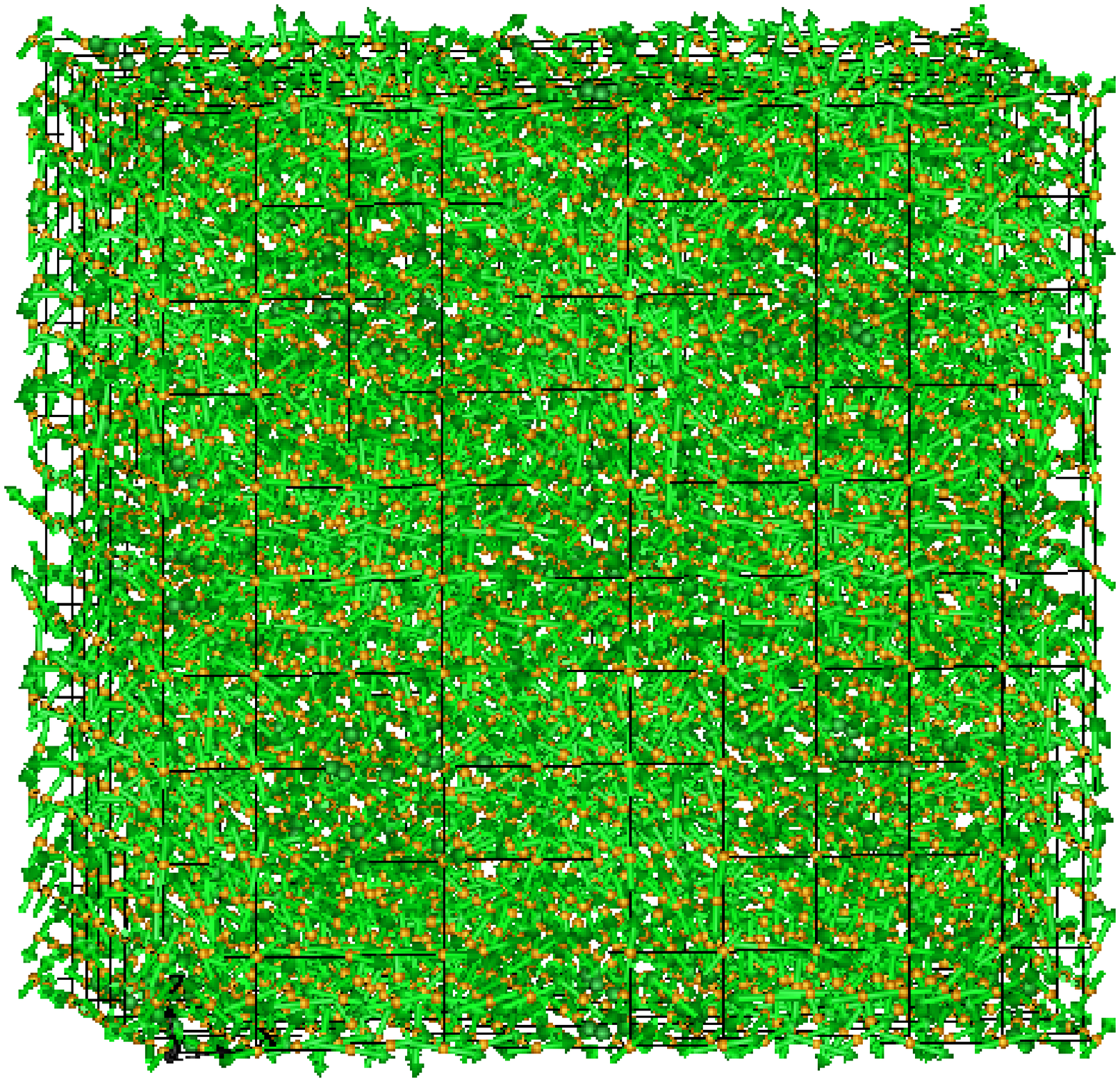} 
\vspace{-0.8cm} \\
(a)
\vspace{5mm} \\
\end{center}
\end{minipage}
\hspace{0.2cm}
\begin{minipage}{5.5cm}
\begin{center}
\vspace*{3.2cm}
\includegraphics[width=5.5cm]{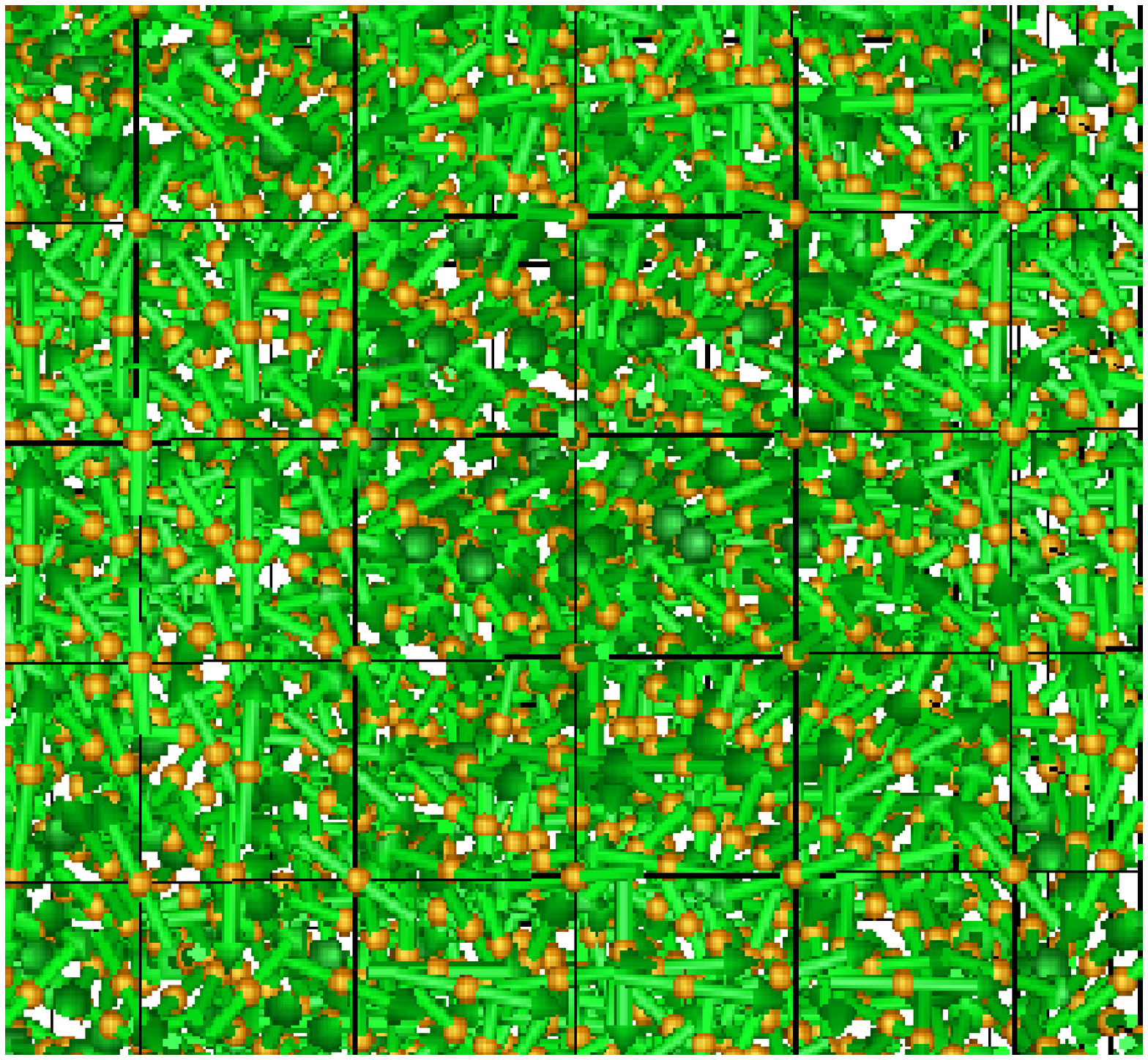} 
\vspace{-0.0cm} \\
(b) 
\vspace{5mm} \\
\end{center}
\end{minipage}
\end{center}
\vspace{0cm}
\caption{(a) Self-consistent AF-base 3QH ($q=0.9$) at $(n,U)=(0.7,9)$. \ \ (b) Enlarged view for showing a local spin structure.  The AF-base vortex structures with size $\lambda = 5a$ are observed on the $x$-$z$ plane.  No helicity is observed in these vorticies.  See also Figs. \ref{fg3qhinb} and \ref{fg3qh} for comparison with the 3QH.  Use a zoom-in tool to see more details of structure.
}
\label{fgaf3qh}
\end{figure}

 The AF-base 3QH structures with $q=0.9$ are stabilized at $(n,U)=(0.7, 9)$, $(0.7, 8.5)$, and $(0.6, 8)$.  Figure \ref{fgaf3qh} shows the self-consistent 3QH structure at (0.7, 9) in the real space.  We observe the magnetic particle structure with size $\lambda = 5a$ on the $x$-$z$ plane, which is similar to the 3QH in Fig. \ref{fg3qh}.  Each particle however hardly shows a net polarization here because neighboring spin configuration is almost antiferromagnetic.
%
%
%
%

\begin{figure}[htbp]
\begin{center}
\includegraphics[width=9cm]{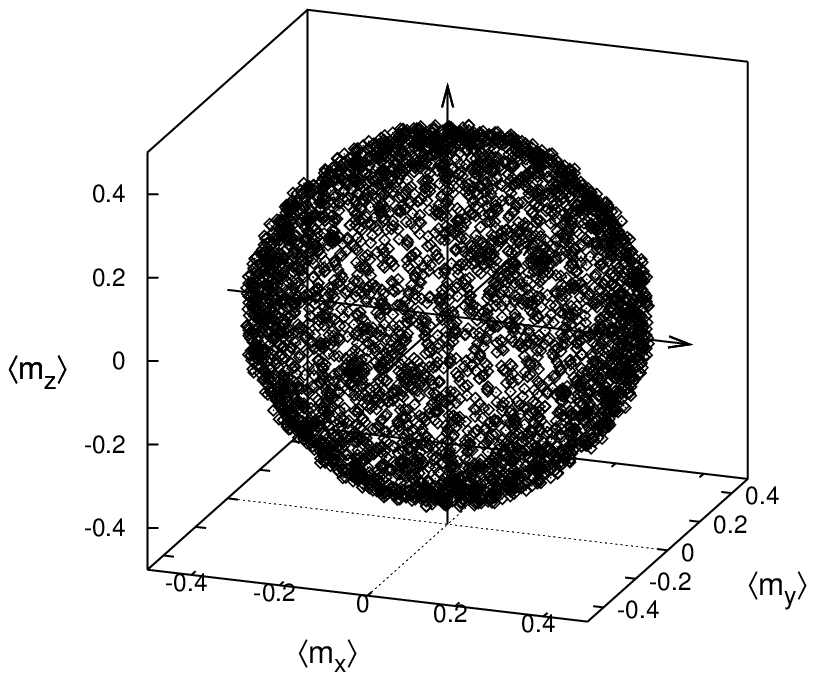}
\end{center}
\vspace{0cm}
\caption{Local moment distribution for the self-consistent AF-base 3QH ($q=0.9$) at $(n,U)=(0.7,9)$.
}
\label{fgaf3qhlm}
\end{figure}

 The Fourier analysis verifies that the AF-base 3QH is given by Eq. (\ref{3qm}) with amplitude $|\bm(\bQ)| = 0.095$ and $q=0.9$, though it is accompanied by 12$Q$ multiple helical SDW (12QMHSDW).  The first 12QMHSDW with amplitude $|\bm(\bQ)| = 0.064$ consists of the wave vectors $\bQ = (q, \pm q', 0)$, $(q, 0, \pm q')$,  $(\pm q', q, 0)$, $(0, q, \pm q')$,  $(\pm q', 0, q)$, and 
 $(0, \pm q', q)$ with $q'=0.2$.  The second 12QMHSDW with $|\bm(\bQ)| = 0.017$ has the same form but $q'=0.4$.  The LM distribution shows a bumpy sphere form due to small amplitude fluctuations and is nearly uniform in direction as shown in Fig. \ref{fgaf3qhlm}.

%
%
%
%
 
We suggested that the same type of AF-base 3QH is possible at $(n,U)=(0.99, 8)$ in our last paper~\cite{kake18}.  The AF-base 3QH, however, disappears in the present calculations.  In our previous work, we first searched the stable 1QH varying the wave number $q$ for a given set of $(n, U(=8))$ and found $q^{\ast} (=0.9)$ for the minimum energy of 1QH.  Then, we compared the 3QH with the same $q^{\ast}$ with the 1QH as well as the 2QH in energy varying electron number $n$ to be consistent with the phenomenological GL phase diagram  for given $\bQ$ vectors, and obtained a possible 3QH at $(n, U)=(0.99, 8)$.  
In the present work, we calculated the energies of the 1QH, 2QH, and 3QH as well as the conical (C) states varying $q$ for a given $(n, U)$, and determined the stable state.  This yields the 3$Q$ multiple transverse SDW (3QMTSDW) with  
$\hat{\bQ}_{1} = (1, 0, 0)$, $\hat{\bQ}_{2} = (0, 1, 0)$, and $\hat{\bQ}_{3} = (0, 0, 1)$ at $(0.99, 8)$.
%
%
\begin{figure}[htbp]
\begin{center}
\includegraphics[width=12cm]{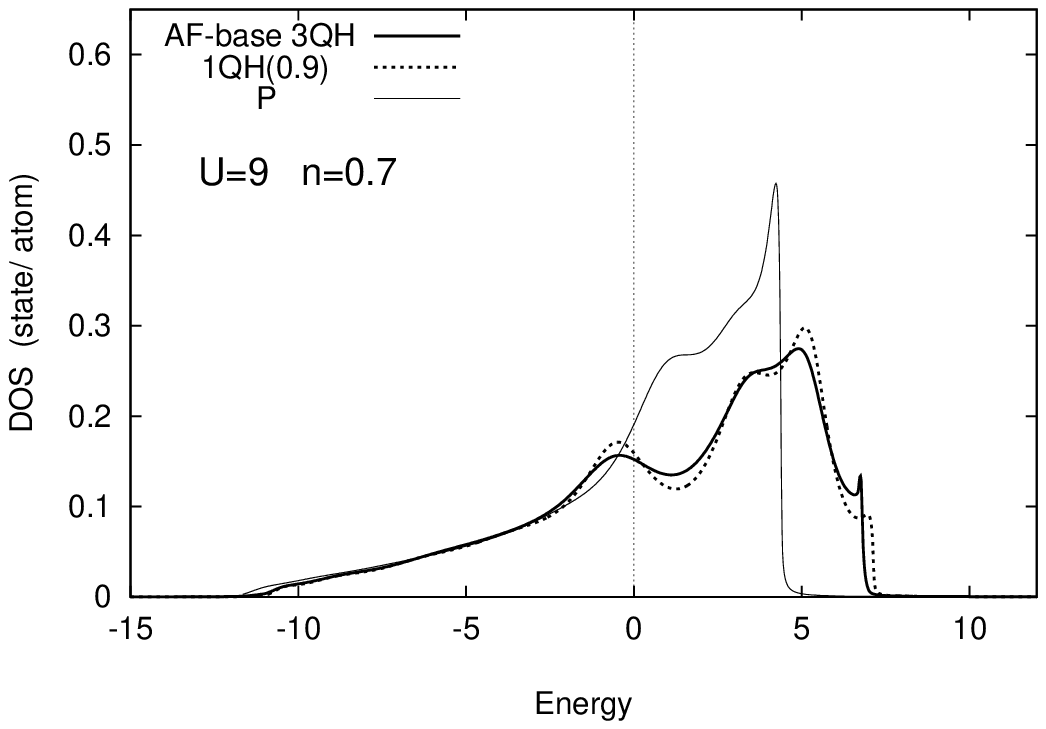}
\end{center}
\vspace{0cm}
\caption{DOS for the self-consistent AF-base 3QH ($q=0.9$), 1QH ($q=0.9$), and the paramagnetic state (P) at $(n,U)=(0.7,9)$.
}
\label{fgdsaf3qh}
\end{figure}
%
%

 The DOS for the AF-base 3QH at $(n, U)=(0.7, 9)$ are shown in Fig. \ref{fgdsaf3qh}.  The AF-base 3QH is regarded as a frustrated MSDW because of no dip near the Fermi level.  Although the AF-base 3QH DOS has a broad peak near the Fermi level, the 1QH DOS forms larger peak in the same energy region, and the paramagnetic state (P) has larger DOS at the Fermi level, so that we find the kinetic energy gain when the AF-base 3QH is compared with these states.  This energy gain stabilizes the AF-base 3QH structure.

\subsection{4$Q$ multiple SDW}

We obtained the Multiple SDW consisting of 4$Q$ wave vectors (4QMSDW) around $(n, U)=(1.4, 5)$, though we started from the 3QH input magnetic structure in our self-consistent calculations.  The 4QMSDW obtained in the range $4.5 \le U \le 6.5$ are the superposition of the 4 Multiple Longitudinal SDW (4QMLSDW).
\begin{align}
\langle \bm_{l} \rangle &= 
m \big[ \be_{1} \cos (\bQ_{1} \! \cdot \! \boldsymbol{R}_{l} + \frac{\pi}{4}) +
\be_{2} \cos (\bQ_{2} \! \cdot \! \boldsymbol{R}_{l} + \frac{\pi}{4})  \nonumber \\
&\ \ \ \ +
\be_{3} \cos (\bQ_{3} \! \cdot \! \boldsymbol{R}_{l} + \frac{\pi}{4}) + 
\be_{4} \cos (\bQ_{4} \! \cdot \! \boldsymbol{R}_{l} + \frac{\pi}{4}) \big] \ .
\label{4qlsdw}
\end{align}
Here 4 wave vectors are given by $\bQ_{1} = (q, q, q)$, $\bQ_{2} = (q, -q, -q)$, 
$\bQ_{3} = (-q, q, -q)$, and $\bQ_{4} = (-q, -q, q)$, respectively, and 4 polarization unit vectors $\be_{i}$ are parallel to $\bQ_{i}$, respectively.  The 4QMSDW with the same $4\bQ$ vectors has recently been found in MnSi${}_{1-x}$Ge${}_{x}$ ($0.25 < x < 0.7$) alloys~\cite{fuji19}. 
%
%
\begin{figure}[htbp]
\begin{center}
\begin{minipage}{10.0cm}
\begin{center}
\includegraphics[width=10.0cm]{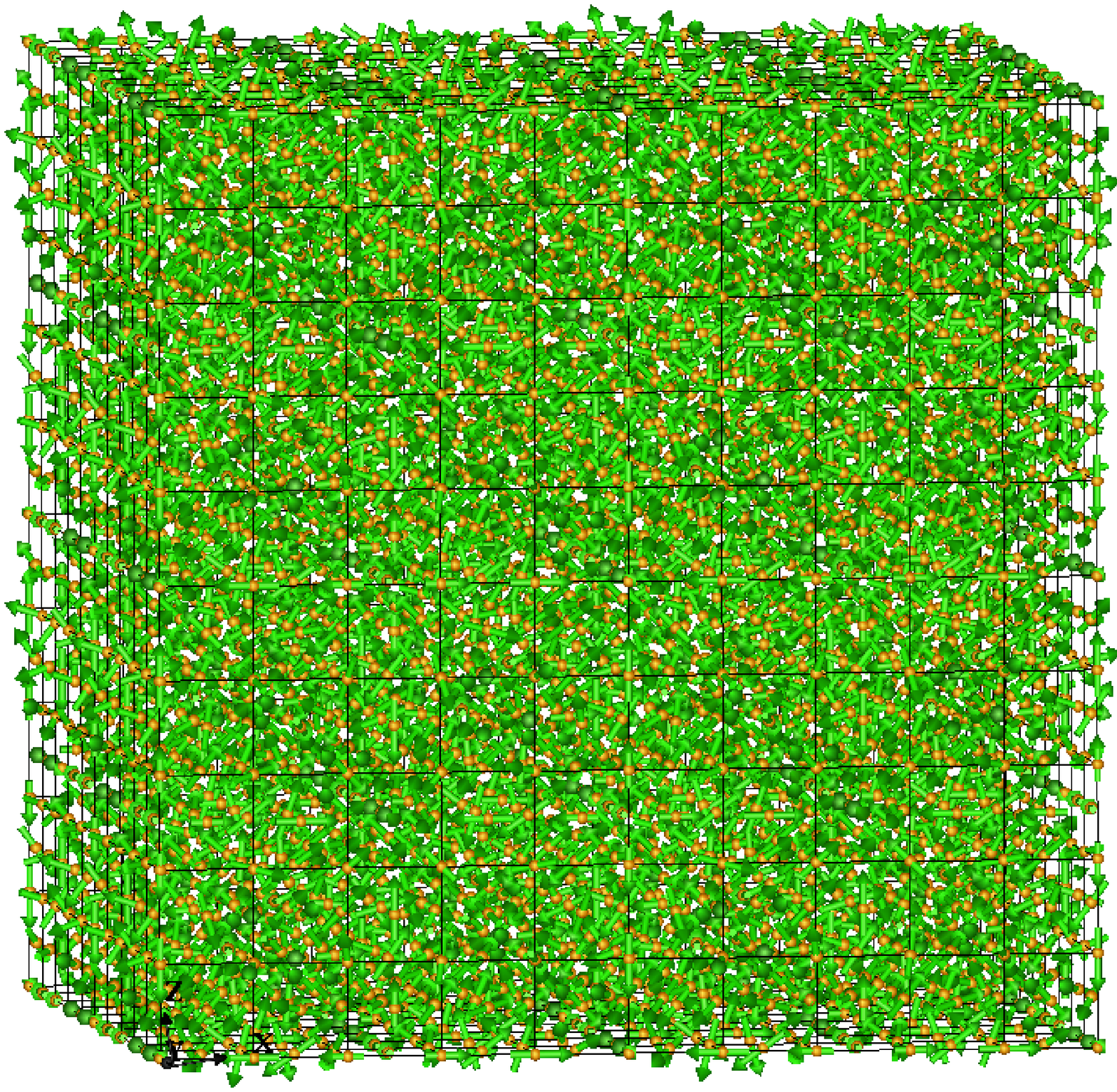} 
\vspace{-0.5cm} \\
(a)
\vspace{5mm} \\
\end{center}
\end{minipage}
\hspace{0.2cm}
\begin{minipage}{5cm}
\begin{center}
\vspace*{4.2cm}
\includegraphics[width=5cm]{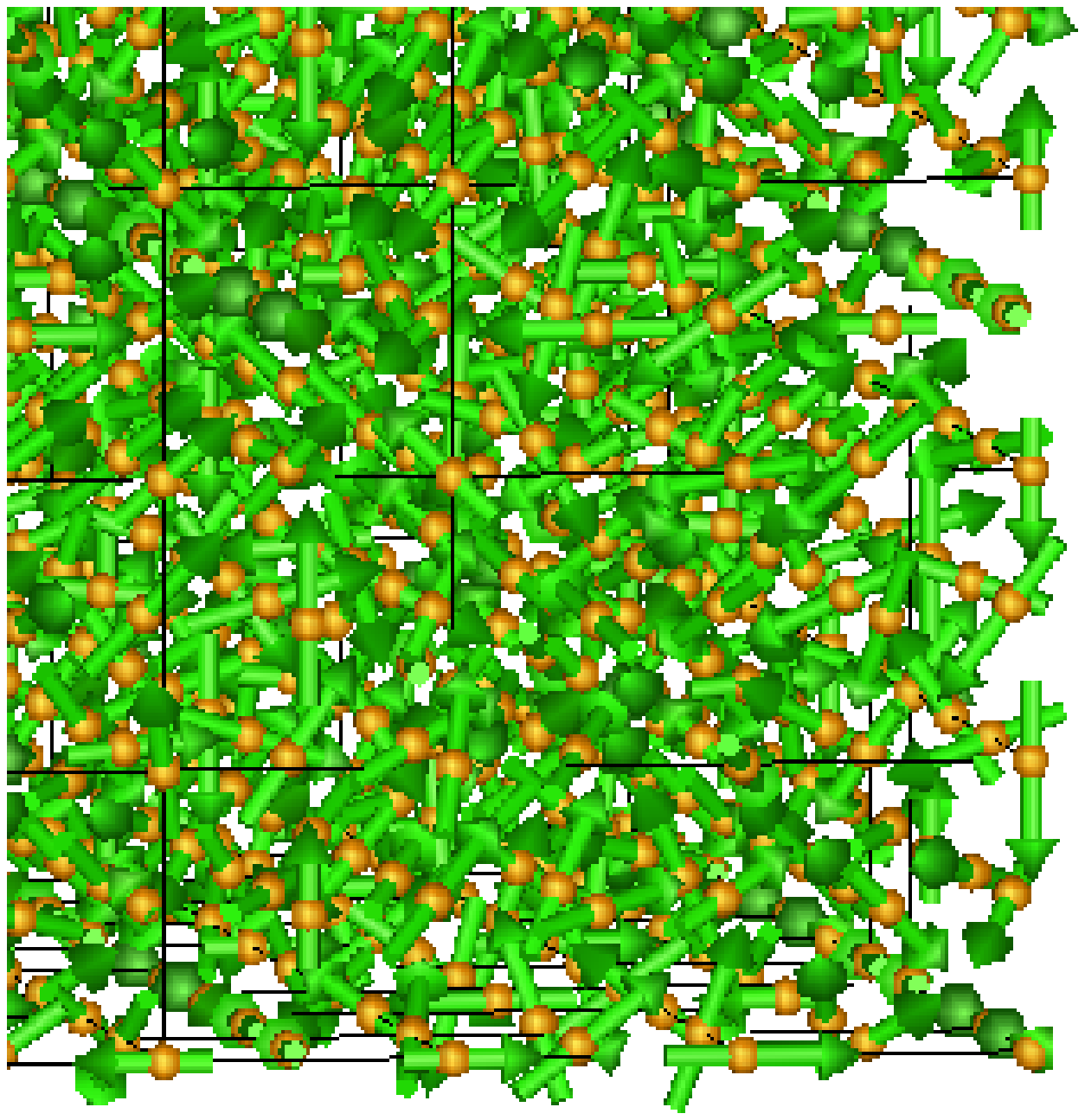} 
\vspace{-0.3cm} \\
(b) 
\vspace{5mm} \\
\end{center}
\end{minipage}
\end{center}
\vspace{0cm}
\caption{(a) Self-consistent 4QMLSDW ($q=0.2$) structure obtained at $(n,U)=(1.7,5)$. \ \ (b) Enlarged view for showing a local spin structure.  Use a zoom-in tool to see more detailed structure.
}
\label{fg4qm}
\end{figure}
%
%

\begin{figure}[htbp]
\begin{center}
\begin{minipage}{7cm}
\begin{center}
\vspace*{-1.0cm}
\includegraphics[width=7cm]{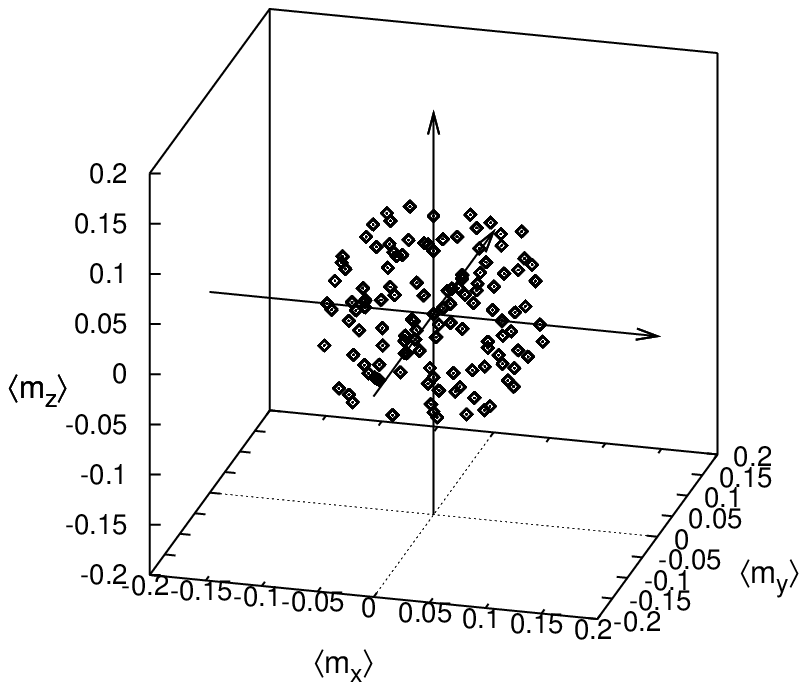} 
\vspace{-1.5cm} \\
(a) 
\vspace{5mm} \\
\end{center}
\end{minipage}
\hspace{0.7cm}
\begin{minipage}{7cm}
\begin{center}
\vspace*{0cm}
\includegraphics[width=7cm]{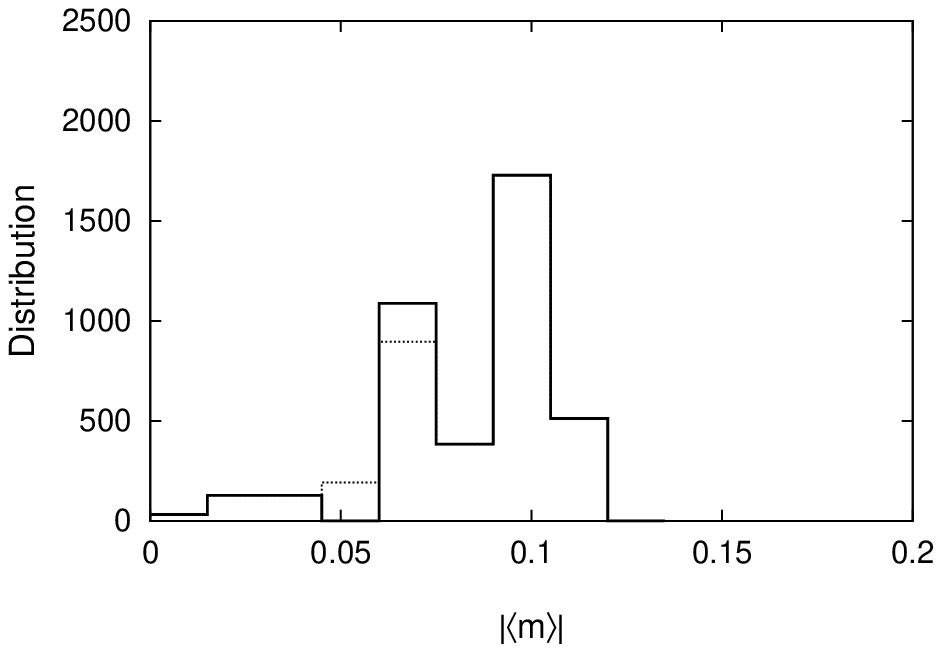} 
\vspace{-0.5cm} \\
(b)
\vspace{5mm} \\
\end{center}
\end{minipage}
\end{center}
\vspace{0cm}
\caption{(a) Local moment distribution for the self-consistent 4QMLSDW ($q=0.2$) at $(n,U)=(1.7,5)$ \ \  (b) Amplitude distribution of local moments for the same 4QMLSDW. Dotted histogram is the distribution for the ideal 4QMLSDW.
}
\label{fg4qmlm}
\end{figure}

The 4QMLSDW with $q=0.2$ at $(n, U)=(1.7, 5)$ and $(1.6, 5)$ do not show any vortex structure as shown in Fig. \ref{fg4qm}.  The LM distribution has a cubic form since the magnetic structure is constructed by the 4 polarization vectors $\{ \be_{i} \}$ (see Fig. \ref{fg4qmlm}).  Accordingly, we find a broad amplitude distribution as shown in Fig. \ref{fg4qmlm}.  A remarkable point is that the magnetic moments with zero amplitude, {\it i.e.}, the non-magnetic atoms appear as the result of the  superposition of the 4 longitudinal waves with the same phase (see, for example, the LM at the corners of the fcc cluster in Fig. \ref{fg4qm}).

%
%
%
%

\begin{figure}[htbp]
\begin{center}
\begin{minipage}{6cm}
\begin{center}
\vspace*{0.5cm}
\includegraphics[width=6cm]{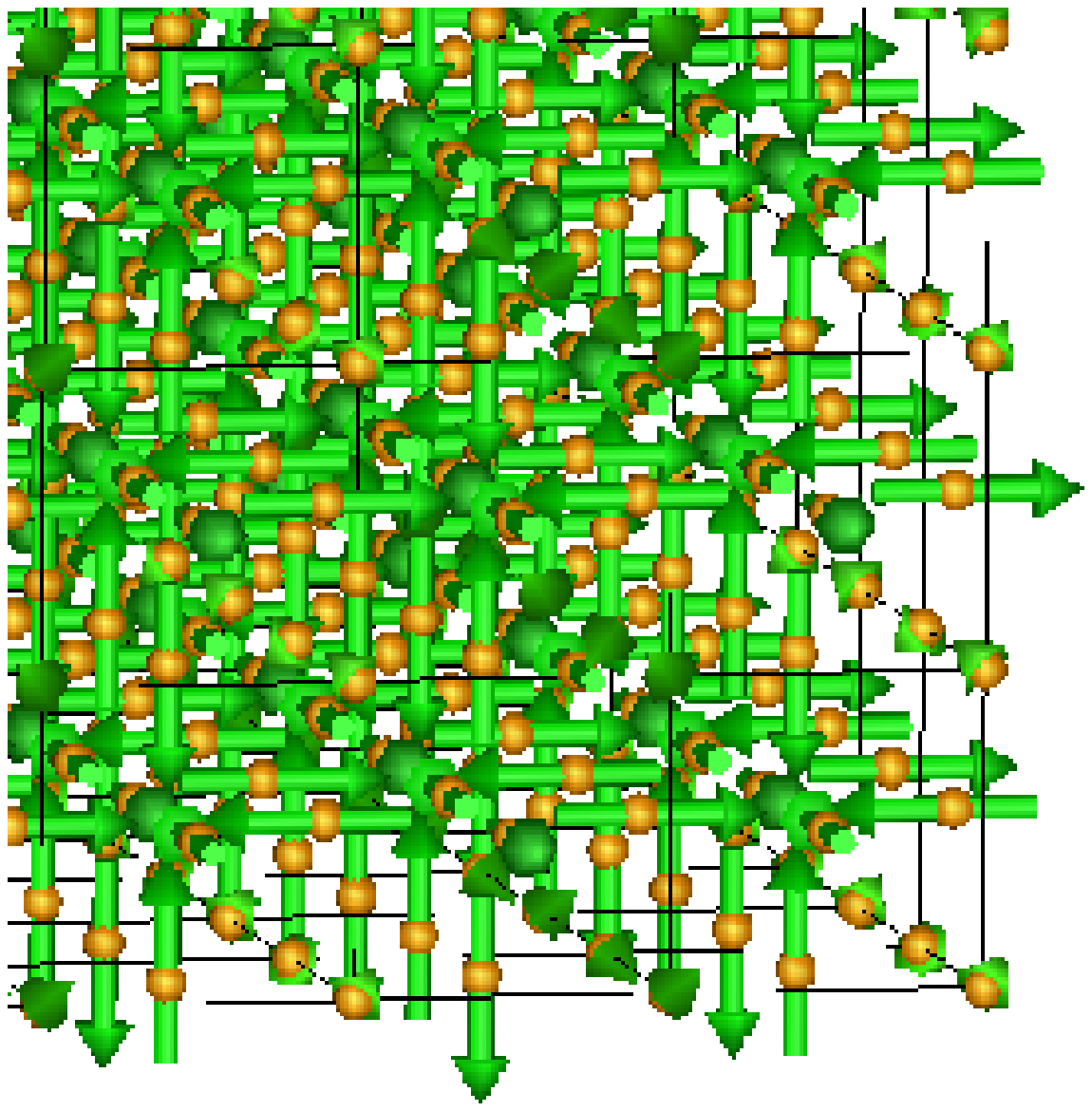} 
\vspace{-0.5cm} \\
(a)
\vspace{5mm} \\
\end{center}
\end{minipage}
\hspace{0.7cm}
\begin{minipage}{8cm}
\begin{center}
\vspace*{-0.4cm}
\includegraphics[width=8cm]{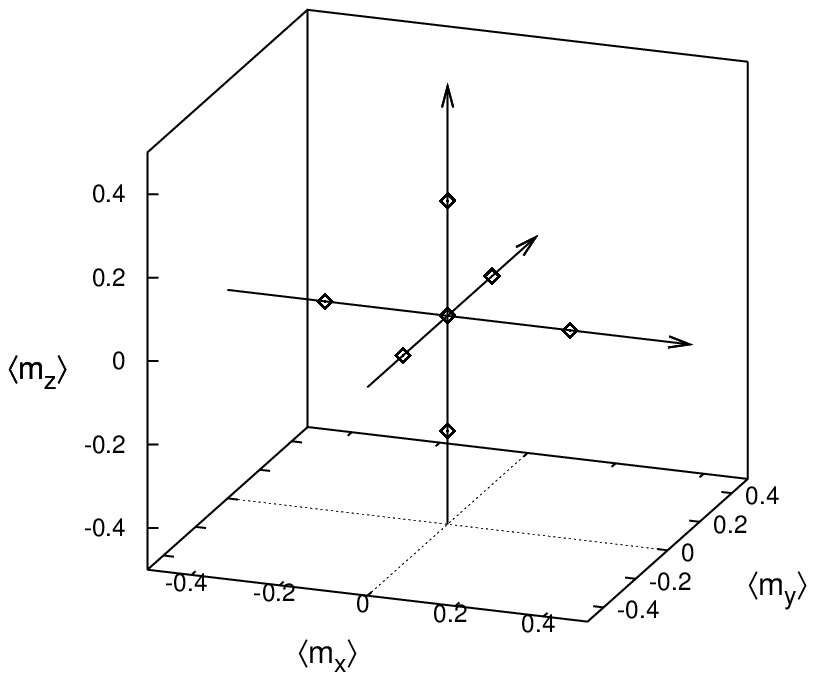} 
\vspace{-1.3cm} \\
(b) 
\vspace{5mm} \\
\end{center}
\end{minipage}
\end{center}
\vspace{0cm}
\caption{Self-consistent 4QMLSDW ($q=0.5$) at $(n,U)=(1.4,5)$. \ \ (a) Real-space structure showing ``all-in'' and ``all-out'' octahedron network (enlarged). \ \ (b) Local-moment distribution showing the appearance of the nonmagnetic sites.
}
\label{fg4qm05}
\end{figure}

When the electron number $n$ is decreased, the wave number $q$ of the 4QMLSDW increases.  We find $q=0.3$ at $(n, U)=(1.55, 6)$, $q=0.4$ at $(1.5, 4.5)$, $(1.5, 5)$, $(1.5, 6)$, $(1.4, 5.5)$, $(1.3, 5.5)$, and $(1.3, 7.5)$, and $q=0.5$ at $(1.4, 4.5)$, $(1.4, 5)$, $(1.3,4.5)$, $(1.3, 5)$, and $(1.2, 5)$, respectively.
In particular, the 4QMLSDW with $q=0.5$ shows rather simple magnetic structure as shown in Fig. \ref{fg4qm05}.  The atoms at the corners of a fcc cubic unit cell are nonmagnetic in this structure.  Remaining 6 face-centered atoms form an octahedron in each unit cell.  The LM's on each octahedron either point to the body center or point in the opposite direction to the center.  Two types of the ``all-in'' and ``all-out'' octahedron magnetic structures are arranged alternatively.
%
%
%
%

\begin{figure}[htbp]
\begin{center}
\includegraphics[width=12cm]{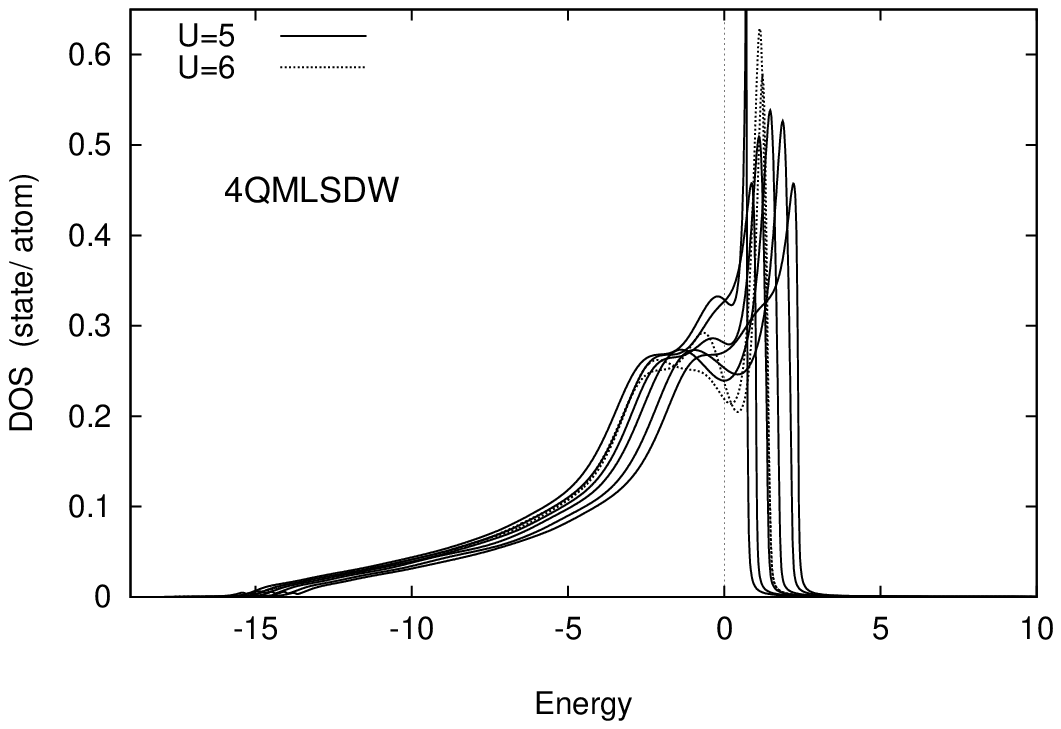}
\end{center}
\vspace{0cm}
\caption{DOS for various self-consistent 4QMLSDW.  Solid curves: the DOS for $U=5$ ($n=1.2$, $1.3$, $1.4$, $1.5$, $1.6$, and $1.7$), dotted curves: the DOS for $U=6$ ($n=1.5$ and $1.55$).
}
\label{fg4qmds}
\end{figure}

The DOS for the 4QMLSDW are presented in Fig. \ref{fg4qmds}.  We find again a dip in the DOS near the Fermi level, being common to these MSDW.  Associated kinetic-energy gain stabilizes the 4QMLSDW.
The magnetic structures with nonmagnetic sites are known as the ``partially ordered state''~\cite{diep13,iked03}.  Although the latter is usually discussed from the viewpoint of the short-range spin fluctuations or frustrations in the local-moment system,  the ``partially ordered state'' here is realized as a superposition of 4 longitudinal SDW, and is stabilized by the band energy gain of the MSDW with well defined dip of DOS in the vicinity of the Fermi level.
%
%
%
%

The 4QMLSDW are not stable for larger $U$ because the strong amplitude modulations accompanied by the structure cause the loss of Coulomb energy.  We found the 4$Q$ Multiple Helical SDW (4QMHSDW) for $(n, U)=(1.45, 7)$ and $(1.47, 7)$.  The principal term of this state is expressed by
\begin{align}
\langle \bm_{l} \rangle &= m_{1} ( \be_{11} \cos \bQ_{1} \! \cdot \! \boldsymbol{R}_{l} + 
\be_{12} \sin \bQ_{1} \! \cdot \! \boldsymbol{R}_{l} )  \nonumber \\
&\ \ \ +
m_{2} ( \be_{21} \cos \bQ_{2} \! \cdot \! \boldsymbol{R}_{l} + 
\be_{22} \sin \bQ_{2} \! \cdot \! \boldsymbol{R}_{l} )  \nonumber \\
&\ \ \ +
m_{3} ( \be_{31} \cos \bQ_{3} \! \cdot \! \boldsymbol{R}_{l} + 
\be_{32} \sin \bQ_{3} \! \cdot \! \boldsymbol{R}_{l} )  \nonumber \\
&\ \ \ +
m_{4} ( \be_{41} \cos \bQ_{4} \! \cdot \! \boldsymbol{R}_{l} + 
\be_{42} \sin \bQ_{4} \! \cdot \! \boldsymbol{R}_{l} )  \ .
\label{4qh}
\end{align}
Here $m_{n} \, (n=1 \sim 4)$ are the amplitudes and the wave vectors $\bQ_{1} \sim \bQ_{4}$ are the same as found in the 4QMLSDW (see Eq. (\ref{4qlsdw})), and $q=0.4$ in the present case.
The polarization unit vectors have some relations; $\be_{31} \approx \be_{11}$,  $\be_{32} \approx -\be_{12}$, $\be_{41} \approx -\be_{21}$, and $\be_{42} \approx \be_{22}$ for $(n, U) = (1.45, 7)$, and $\be_{31} \approx \be_{11}$,  $\be_{22} \approx -\be_{12}$, $\be_{41} \approx -\be_{31}$, and $\be_{42} \approx \be_{32}$ for $(n, U) = (1.47, 7)$.  Moreover, $\be_{n1}$ and $\be_{n2}$ ($n=1 \sim 4$) are approximately orthogonal each other.  

We also found the 4QMHSDW at $(n, U) = (0.6, 8.5)$ near the paramagnetic phase boundary.  This structure consists of 4 elliptical helical states whose rotational planes contain the $z$ axis.
\begin{align}
\langle \bm_{l} \rangle &= - m_{1} \be_{1} \cos \bQ_{1} \! \cdot \! \boldsymbol{R}_{l} - 
m_{2} \boldsymbol{k} \sin \bQ_{1} \! \cdot \! \boldsymbol{R}_{l}  \nonumber \\
&\ \ \ +
m_{1} \be_{2} \cos \bQ_{2} \! \cdot \! \boldsymbol{R}_{l} - 
m_{2} \boldsymbol{k} \sin \bQ_{2} \! \cdot \! \boldsymbol{R}_{l}  \nonumber \\
&\ \ \ -
m_{2} \boldsymbol{k} \cos \bQ_{3} \! \cdot \! \boldsymbol{R}_{l} + 
m_{1} \be_{3} \sin \bQ_{3} \! \cdot \! \boldsymbol{R}_{l}  \nonumber \\
&\ \ \ -
m_{2} \boldsymbol{k} \cos \bQ_{4} \! \cdot \! \boldsymbol{R}_{l} - 
m_{1} \be_{4} \sin \bQ_{4} \! \cdot \! \boldsymbol{R}_{l}  \ .
\label{4qh2}
\end{align}
Here $m_{1} = 0.037$ and $m_{2} = 0.021$. The wave vectors $\{  \bQ_{n} \}$ are given by $\bQ_{1} = (q, q', 0)$, $\bQ_{2} = (q, -q', 0)$, $\bQ_{3} = (q', q, 0)$, and $\bQ_{4} = (q', -q, 0)$ with $q=0.9$ and $q'=0.4$.  Note that $\{  \bQ_{n} \}$ are on the $x$-$y$ plane, and the polarization vectors $\be_{1}$, $\be_{2}$, $\be_{3}$, and $\be_{4}$ are approximately parallel to $\bQ_{2}$, $\bQ_{1}$, $\bQ_{4}$, and $\bQ_{3}$, respectively.
 
The 4QMHSDW do not show any vortex-type structure.  The LM distributions are approximately spherical and show the weak amplitude modulations as mentioned above.

\subsection{12$Q$ multiple SDW}

We found the 12$Q$ Multiple SDW (12QMSDW) around the 4QMLSDW region.  These structures are given by
\begin{align}
\langle \bm_{l} \rangle &= 
m \boldsymbol{i} \big[ -\cos (\bQ_{1} \! \cdot \! \boldsymbol{R}_{l} - \frac{\pi}{4}) +
\cos (\bQ_{2} \! \cdot \! \boldsymbol{R}_{l} - \frac{\pi}{4})  \nonumber \\
&\ \ \ \ \ \ \ \ \ +
\cos (\bQ_{3} \! \cdot \! \boldsymbol{R}_{l} - \frac{\pi}{4}) - 
\cos (\bQ_{4} \! \cdot \! \boldsymbol{R}_{l} - \frac{\pi}{4}) \big] \nonumber \\
& +
m \boldsymbol{j} \big[ -\cos (\bQ_{5} \! \cdot \! \boldsymbol{R}_{l} - \frac{\pi}{4}) -
\cos (\bQ_{6} \! \cdot \! \boldsymbol{R}_{l} - \frac{\pi}{4})  \nonumber \\
&\ \ \ \ \ \ \ \ \ +
\cos (\bQ_{7} \! \cdot \! \boldsymbol{R}_{l} - \frac{\pi}{4}) + 
\cos (\bQ_{8} \! \cdot \! \boldsymbol{R}_{l} - \frac{\pi}{4}) \big] \nonumber \\
& +
m \boldsymbol{k} \big[ -\cos (\bQ_{9} \! \cdot \! \boldsymbol{R}_{l} - \frac{\pi}{4}) +
\cos (\bQ_{10} \! \cdot \! \boldsymbol{R}_{l} - \frac{\pi}{4})  \nonumber \\
&\ \ \ \ \ \ \ \ \ -
\cos (\bQ_{11} \! \cdot \! \boldsymbol{R}_{l} - \frac{\pi}{4}) + 
\cos (\bQ_{12} \! \cdot \! \boldsymbol{R}_{l} - \frac{\pi}{4}) \big] \ .
\label{12qsdw}
\end{align}
Here $\bQ_{1} = (q, q', q')$, $\bQ_{2} = (q, -q', -q')$, 
$\bQ_{3} = (-q, q', -q')$, $\bQ_{4} = (-q, -q', q')$, 
$\bQ_{5} = (q', q, q')$, $\bQ_{6} = (q', -q, -q')$, 
$\bQ_{7} = (-q', q, -q')$, $\bQ_{8} = (-q', -q, q')$, 
$\bQ_{9} = (q', q', q)$, $\bQ_{10} = (q', -q', -q)$, 
$\bQ_{11} = (-q', q', -q)$, and $\bQ_{12} = (-q', -q', q)$. 
%
%
\begin{figure}[htbp]
\begin{center}
\begin{minipage}{10.0cm}
\begin{center}
\includegraphics[width=10.0cm]{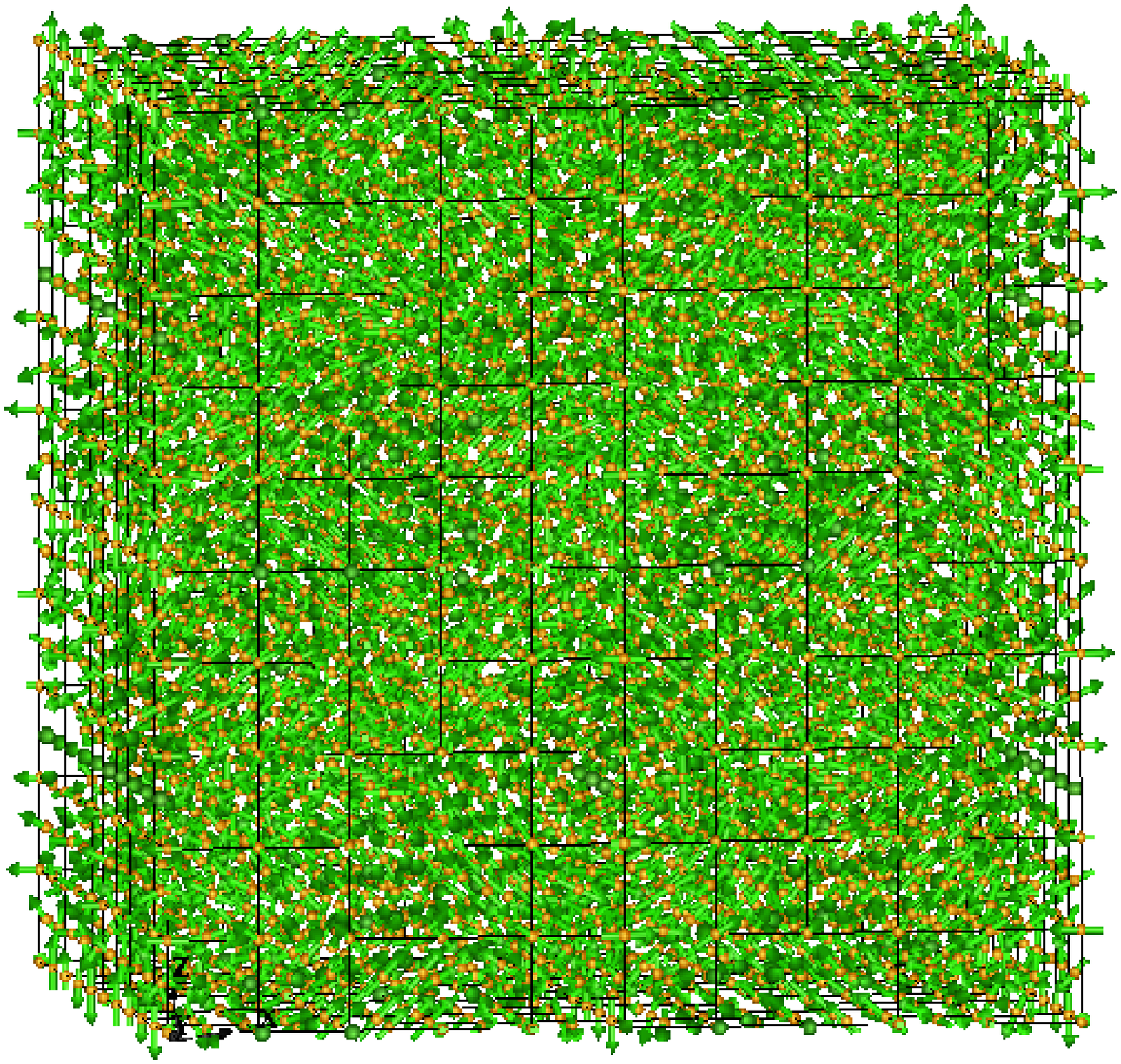} 
\vspace{-0.5cm} \\
(a)
\vspace{5mm} \\
\end{center}
\end{minipage}
\hspace{0.2cm}
\begin{minipage}{5cm}
\begin{center}
\vspace*{4.2cm}
\includegraphics[width=5cm]{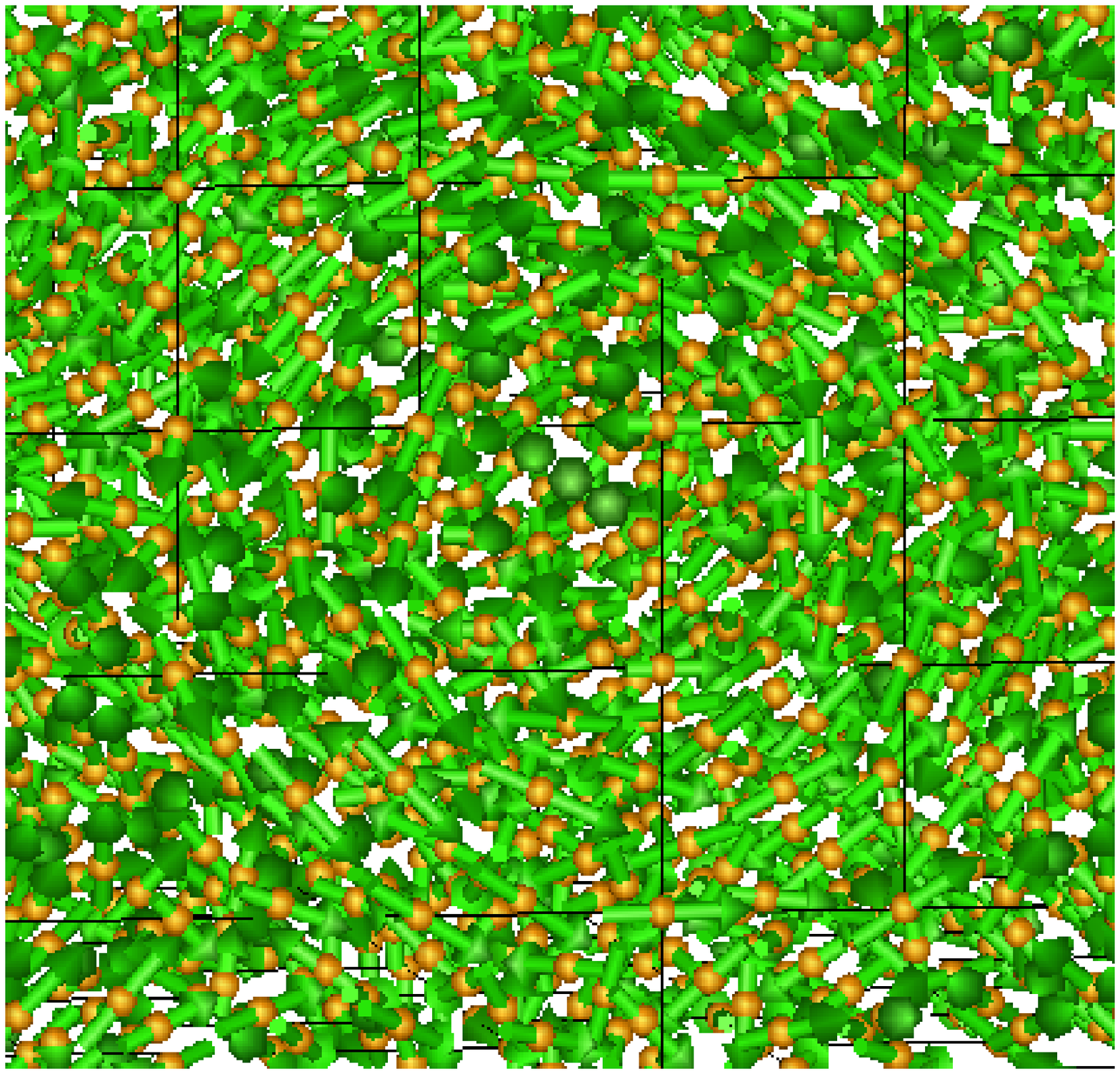} 
\vspace{0.0cm} \\
(b) 
\vspace{5mm} \\
\end{center}
\end{minipage}
\end{center}
\vspace{0cm}
\caption{(a) Self-consistent 12QMSDW ($q=0.1$, $q'=0.2$) structure obtained at $(n,U)=(1.75, 5)$. \ \ (b) Enlarged view for showing a half-skyrmion-type vortex spin structure.  An anticlockwise vortex structure (half-antiskyrmion) is observed at the lower center on the $x$-$z$ plane.  Use a zoom-in tool to see more detailed magnetic structure.
}
\label{fg12qm}
\end{figure}
%
%

\begin{figure}[htbp]
\begin{center}
\begin{minipage}{7cm}
\begin{center}
\vspace*{-1.0cm}
\includegraphics[width=7cm]{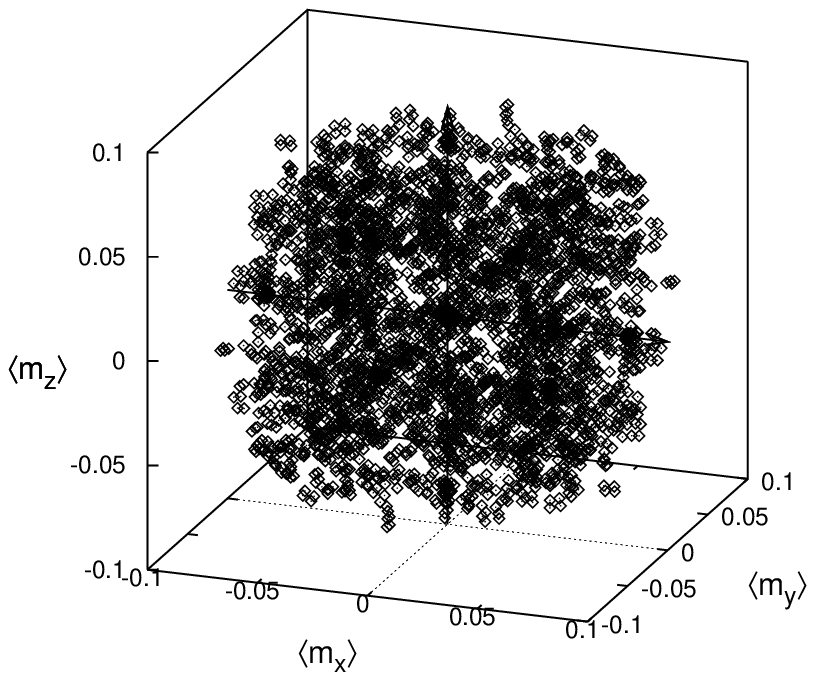} 
\vspace{-1.5cm} \\
(a) 
\vspace{5mm} \\
\end{center}
\end{minipage}
\hspace{0.7cm}
\begin{minipage}{7cm}
\begin{center}
\vspace*{0cm}
\includegraphics[width=7cm]{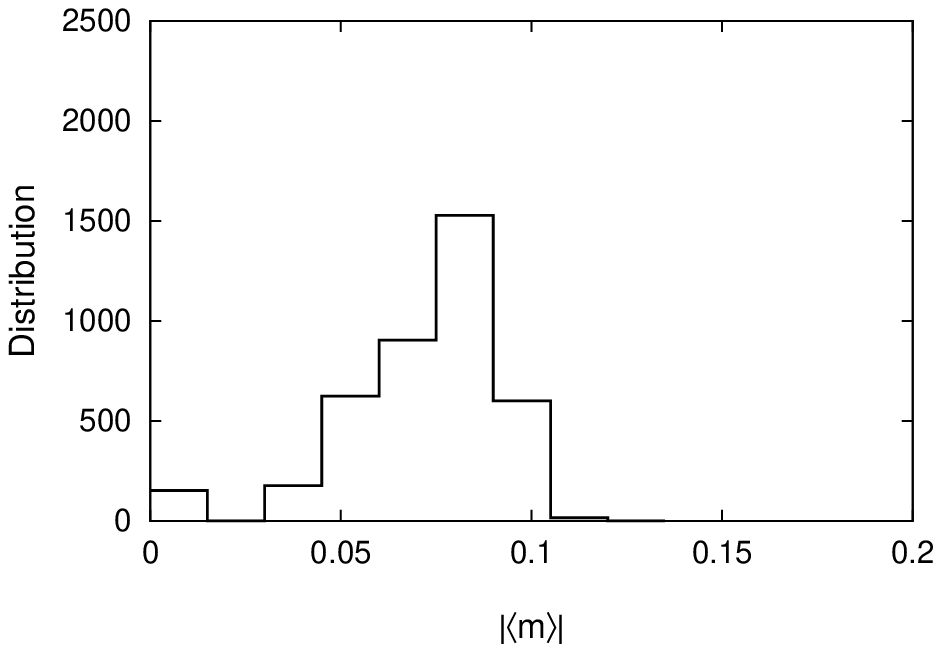} 
\vspace{-0.5cm} \\
(b)
\vspace{5mm} \\
\end{center}
\end{minipage}
\end{center}
\vspace{0cm}
\caption{(a) Local moment distribution for the self-consistent 12QMSDW ($q=0.1$, $q'=0.2$) at $(n,U)=(1.75,5)$ \ \  (b) Amplitude distribution of local moments for the same 12QMSDW.
}
\label{fg12qmlm}
\end{figure}

The 12QMSDW with $q=0.1$ and $q'=0.2$ found at $(n, U)=(1.75 \sim 1.85, 4.75)$, $(1.73, 5)$, $(1.75, 5)$, and $(1.7, 5.25)$ shows a half-skyrmion type vortex structure as seen in Fig. \ref{fg12qm}.  We find the anticlock-wise vortex at the lower center on the $x$-$z$ plane, and halves of the clock-wise vortex at the upper-left and upper-right corners, respectively.  The LM distribution is the cubic type and the amplitudes show a broad distribution as seen in Fig. \ref{fg12qmlm}.  It is remarkable that the nonmagnetic atoms appear in the 12QMSDW as in the 4QMLSDW.  Therefore the 12QMSDW with $q=0.1$ and $q'=0.2$ shows the half-skyrmion-type vortex structure with partially ordered state.  
%
%
%
%

\begin{figure}[htbp]
\begin{center}
\includegraphics[width=12cm]{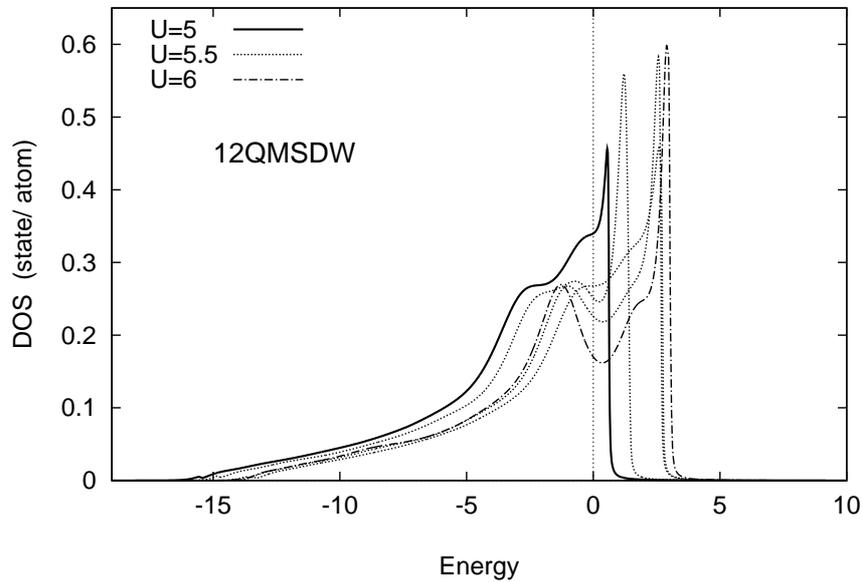}
\end{center}
\vspace{0cm}
\caption{DOS for various self-consistent 12QMSDW.  Solid curve: the DOS for $U=5$ and $n=1.75$,  dotted curves: the DOS for $U=5.5$ ($n=1.1$, $1.2$, and $1.5$), Dot-dashed curve: the DOS for $U=6$ and $n=1.2$.
}
\label{fg12qmds}
\end{figure}

The same type of 12QMSDW with $q=0.3$ and $q'=0.4$, and $q=0.6$ and $q'=0.4$ appear at $(n, U)=(1.5, 5.5)$, $(1.1, 5.5)$, $(1.2, 5.5)$, and $(1.2, 6)$.  These structures, however, do not show the vortex structure because $q$ and $q'$ are comparable to 1 and polarization vectors deviate from $\boldsymbol{i}$, $\boldsymbol{j}$, $\boldsymbol{k}$, considerably.  In addition the LM distributions become more spherical and the amplitude modulations become weaker because of larger $U$.  Calculated DOS for the 12QMSDW again have a dip near the Fermi level as shown in Fig. \ref{fg12qmds}.  The stability of 12QMSDW is therefore attributed to the kinetic energy gain due to the dip formation near the Fermi level.
%
%
%
%

\subsection{Other magnetic structures}

We found many other MSDW in the present calculations as seen in Fig. \ref{fgunpd}.  The magnetic structure at $(n, U)=(1.0, 7)$, for example, is the 2Q-MSDW consisting of a helical SDW and a longitudinal SDW (LSDW).
\begin{align}
\langle \bm_{l} \rangle &= m_{1} (
\boldsymbol{i} \cos \boldsymbol{Q}_{1} \! \cdot \! \boldsymbol{R}_{l} -
\boldsymbol{j} \sin \boldsymbol{Q}_{1} \! \cdot \! \boldsymbol{R}_{l} )  -
m_{2} \boldsymbol{k} \cos \boldsymbol{Q}_{2} \! \cdot \! \boldsymbol{R}_{l}   \ .
\label{hlsdw}
\end{align}
Here $m_{1}=0.28$, $m_{2}=0.21$, $\bQ_{1} = (0, 0, 0.9)$ 
($\bot \ \boldsymbol{i}, \boldsymbol{j}$), and $\bQ_{2} = (0, 0, 0.8)$ 
($/ \! / \ \boldsymbol{k}$).
The ordered state at $(n, U)=(0.9, 8)$ is a simple 2$Q$ Multiple Transverse SDW (2QMTSDW), which is given by
\begin{align}
\langle \bm_{l} \rangle &= m (
\boldsymbol{j} \cos \hat{\bQ}_{1} \! \cdot \! \boldsymbol{R}_{l} +
\boldsymbol{k} \cos \hat{\bQ}_{2} \! \cdot \! \boldsymbol{R}_{l} ) \ ,
\label{2qtsdw}
\end{align}
with $m = 0.46$, $\hat{\bQ}_{1} = (1, 0, 0)$, and $\hat{\bQ}_{2} = (0, 1, 0)$.

Between the above-mentioned 2$Q$-MSDW's, the 3$Q$-MTSDW structures appear.  The 3$Q$-MTSDW at $(n, U)=(0.9, 6.5)$, $(0.8, 7.5)$, $(1.0, 7.5)$, $(1.0, 8)$, $(0.9, 8.5)$, $(1.0, 8.5)$, $(0.9, 9)$, and $(1.0, 9)$ are expressed as
\begin{align}
\langle \bm_{l} \rangle &= m \big[ 
\boldsymbol{j} \cos (\hat{\bQ}_{1} \! \cdot \! \boldsymbol{R}_{l} - \alpha_{1}) +
\boldsymbol{k} \cos (\hat{\bQ}_{2} \! \cdot \! \boldsymbol{R}_{l} - \alpha_{2}) + 
\boldsymbol{i} \cos (\hat{\bQ}_{3} \! \cdot \! \boldsymbol{R}_{l} - \alpha_{3}) \big] \ ,
\label{3qtsdw}
\end{align}
where $\hat{\bQ}_{1} = (1, 0, 0)$, $\hat{\bQ}_{2} = (0, 1, 0)$, $\hat{\bQ}_{3} = (0, 0, 1)$, and 
$\alpha_{1}=\alpha_{2}=\alpha_{3}=0$.  The other 3Q-MTSDW at $(n, U)=(0.8, 7)$ and
$(0.9, 7)$ have the same form, but take different phases $\alpha_{1}$, $\alpha_{2}$, and $\alpha_{3}$.

For weaker Coulomb interaction $U=6.5$ and $n=1.0$, we found the 3$Q$ Multiple Longitudinal SDW (3QMLSDW) given by
\begin{align}
\langle \bm_{l} \rangle = -m \, ( 
\boldsymbol{i} \sin \bQ_{1} \! \cdot \! \boldsymbol{R}_{l} +
\boldsymbol{j} \sin \bQ_{2} \! \cdot \! \boldsymbol{R}_{l} + 
\boldsymbol{k} \sin \bQ_{3} \! \cdot \! \boldsymbol{R}_{l} \, ) \ ,
\label{3qlsdw}
\end{align}
with $m = 0.23$, $\bQ_{1} = (0.8, 0, 0)$, $\bQ_{2} = (0, 0.8, 0)$, and $\bQ_{3} = (0, 0, 0.8)$.

The ordered state found at $(0.8, 10)$ is an AF-base 3$Q$-MSDW consisting of a TSDW with 
$\hat{\bQ}_{1} = (0, 0, 1)$ and the 2$Q$ multiple helical SDW with $\bQ_{2} = (0.2, 0, 1)$ and 
$\bQ_{3} = (-0.2, 0, 1)$.
\begin{align}
\langle \bm_{l} \rangle &= 
m_{1} \boldsymbol{i} \cos \hat{\bQ}_{1} \! \cdot \! \boldsymbol{R}_{l}  \nonumber \\
& + m_{2} \,  (
\boldsymbol{k} \cos \boldsymbol{Q}_{2} \! \cdot \! \boldsymbol{R}_{l} +
\boldsymbol{j} \sin \boldsymbol{Q}_{2} \! \cdot \! \boldsymbol{R}_{l}  +
\boldsymbol{k} \cos \boldsymbol{Q}_{3} \! \cdot \! \boldsymbol{R}_{l} -
\boldsymbol{j} \sin \boldsymbol{Q}_{3} \! \cdot \! \boldsymbol{R}_{l} ) \ .
\label{t2qsdw}
\end{align}
Here $m_{1} = 0.56$ and $m_{2} = 0.14$.  The first TSDW creates the AF array on the $y$-$z$ plane and the second 2QMHSDW causes a screw rotation of magnetic moments on the $y$-$z$ plane with the translation along the $x$ axis.

The ordered state found at $(0.7, 10)$ shows more complex AF-base 10Q-MSDW whose principal terms are expressed by
\begin{align}
\langle \bm_{l} \rangle &= \boldsymbol{j} \, \big[ \,
m_{1} \cos \hat{\bQ}_{1} \! \cdot \! \boldsymbol{R}_{l}  \nonumber \\
& - m_{2} \,  (
\cos \boldsymbol{Q}_{2} \! \cdot \! \boldsymbol{R}_{l}  +
\cos \boldsymbol{Q}_{3} \! \cdot \! \boldsymbol{R}_{l} ) 
- m_{3} \,  (
\cos \boldsymbol{Q}_{4} \! \cdot \! \boldsymbol{R}_{l} + 
\cos \boldsymbol{Q}_{5} \! \cdot \! \boldsymbol{R}_{l} ) 
\, \big] \ ,
\label{5qsdw}
\end{align}
where $m_{1}=0.29$, $m_{2}=0.21$, $m_{3}=0.18$, $\hat{\bQ}_{1} = (1, 0, 0)$, $\bQ_{2} = (1, 0.4, 0)$, $\bQ_{3} = (1, -0.4, 0)$, 
$\bQ_{4} = (0, 0.2, 1)$, and $\bQ_{5} = (0, -0.2, 1)$.  Remaining 5 terms consist of a TSDW 
with $\bQ = (0, 0, 1)$ and polarization vector $\boldsymbol{i}$, and 4 elliptical helical SDW on the $x$-$z$ plane.

The magnetic state at $(n, U) = (1.0, 10)$ is an AF-base complex MSDW, which is not described by a superposition of several SDW's, according to the Fourier analysis.  
We finally remark that all the magnetic structures mentioned in this subsection are not vortex type.

\section{Summary}

We have investigated the stability of the vortex-type magnetic structures in itinerant electron system with inversion symmetry on the basis of the Hubbard model and the generalized Hartree-Fock approximation combined with the recursion method.  We determined numerically a possible magnetic phase diagram on the fcc lattice in the weak and intermediate Coulomb interaction region, taking into account various multiple spin density waves (MSDW).

We found the vortex-type magnetic structures with small wave numbers around the Stoner line where the long-wave spin polarization shows an instability.  The 2$Q$ multiple helical structure (2QH) found at $(n, U)=(1.70 \sim 1.73, 5.5)$ and $(1.6, 5.5)$ forms on the (001) plane a giant antiferromagnetic (AF) face-centered square lattice consisting of the half-skyrmion (meron) and half-antiskyrmion (antimeron) vortex particles with size $\lambda = a/q$, $q \, (= 0.2, 0.3 << 1)$ being the wave number.

We found the 3QH vortex structure in the vicinity of the ferromagnetic phase boundary, {\it i.e.},  at $(n, U) = (1.68, 6)$, $(1.6, 6.5)$, and $(1.42, 7.5)$.  The 3QH are regarded as a superposition of the 2QH and the 1QH.  There, the 2QH half-skyrmion type paired vortices with ``westerlies'' on a $(001)$ plane are twisted along $[001]$ direction because of the remaining 1QH wave.  

The 2QH and 3QH vortex-type structures are accompanied by additional 4QMSDW with larger wave numbers.  These additional MSDW reduce the Coulomb energy loss due to amplitude modulations, and suppress their amplitude distributions considerably.

We have clarified the electronic origin of the stability for the 2QH and the 3QH.  The 3QH is stabilized by the kinetic energy gain due to the formation of a sharp dip in the DOS near the Fermi level.  The DOS for the 2QH do not show such a clear dip, but we numerically verified that the kinetic energy gain contributes to the stability when the 2QH are compared with the 1QH in energy.

We found at $(n, U) = (0.7, 9)$, $(0.7, 8.5)$, and $(0.6, 8)$ the AF-base 3QH with the wave number $q=0.9$.  There the half-skyrmion-type vortex particles with size $\lambda = a/2\tilde{q}$ ($\tilde{q}=1-q=0.1$) form a lattice on the $(001)$ plane, but the LM in each vortex particle  alternatively change their direction with translation of half a lattice constant.  This state is also accompanied by the 12Q-MHSDW as satellite contributions.  The AF-base 3QH is a frustrated system which does not show any dip at the Fermi level in the DOS.

We found that the 12QMSDW states are also possible.  The 12QMSDW with the wave numbers $q=0.1$ and $q'=0.2$ shows a clear half-skyrmion and half-antiskyrmion type vortex structures as well as the partially ordered state with zero-moment sites.  The LM distribution shows a cubic form and the amplitudes of LM show a broad distribution, being characteristic of itinerant electron system.  These structures are also stabilized by a dip formation in the DOS near the Fermi level.

Apart from the vortex type structure, we found other MSDW structures.
The 4QMLSDW found in the region of $U=5$ and $5.5$ show a partially ordered state accompanied by nonmagnetic sites.  In particular, the 4QMLSDW ($q=0.5$) at $(n, U) = (1.4,5)$ forms the ``all-in'' and ``all-out'' octahedron network on the face centered sites of the fcc lattice.  The 4QMHSDW and the AF-base 10QMSDW also appear in the stronger $U$ region, at $(1.45, 7)$ and $(0.7, 10)$, respectively.  We found that most of these MSDW, including the 1QH state, are stabilized by the formation of a dip near the Fermi level in the DOS.

Present work indicates that a variety of the multiple spin density waves and vortex-type magnetic structures are possible even in the itinerant electron system with inversion symmetry under zero magnetic field.  It also suggests that the long-range competing interactions being characteristic of the itinerant electron system play an important role in the formations of the magnetic skyrmions found in Co-Zn-Mn and Mn-Si-Ge systems without inversion symmetry~\cite{muhl09,yu18,zhan16,fuji19}.  It should be noted that the size of cluster in the present calculations is not enough to cover larger size of vortex magnetic structures as found in the experiments.  Furthermore the GHF approach does not necessarily guarantee the global minimum of energy for magnetic states.  In order to obtain more solid conclusion, we have to perform calculations increasing the number of input structures, using larger size of cluster, and applying advanced methods which automatically yield the magnetic structure with the global minimum~\cite{kake98,kimu00,uchida16}.  The calculations are left for future work.

\begin{acknowledgment}


The author would like to express his thanks to Prof. T. Uchida
 for valuable discussions and comments on the present work. 

\end{acknowledgment}




\begin{thebibliography}{99}
%
%
\bibitem{skyr61} T.H.R. Skyrme, Proc. R. Soc. A{\bf 260}, 127 (1961); Nucl. Phys. {\bf 31}, 556  (1962).
%
%
\bibitem{bogd89} A.N. Bogdanov and D.A. Yablonskii, Sov. Phys. JETP {\bf 68}, 101 (1989).
%
%
\bibitem{bogd94} A. Bogdanov and A. Hubert, J. Magn. Magn. Mat. {\bf 138}, 255 (1994).
%
%
\bibitem{bogd01} A.N. Bogdanov and U.K. R\"{o}{\ss}ler, Phys. Rev. Lett. {\bf 87}, 037203 (2001).
%
%
\bibitem{ross06} U.K. R\"{o}{\ss}ler, A.N. Bogdanov, and C. Pfleiderer, Nature {\bf 442}, 797 (2006).
%
%
\bibitem{muhl09} S. M\"{u}hlbauer, B. Binz, F. Jonietz, C. Pfleiderer, A. Rosch, A. Neubauer, and R. B\"{o}ni, Science  {\bf 323}, 915 (2009). 
%
%
\bibitem{yu11} X. Z. Yu, N. Kanazawa, Y. Onose, K. Kimoto, W.Z. Zhang, S. Ishiwata, Y. Matsui,  and Y. Tokura, Nat. Mater. {\bf 10}, 106 (2011). 
%
%
\bibitem{fert17} A. Fert, N. Reyren, and V. Cros, Nature Rev. {\bf 2}, 17031 (2017).
%
%
\bibitem{ever18} K. Everschor-Sitte, J. Masell, R.M. Reeve, and M. Kl\"{a}ui, J. Appl. Phys. {\bf 124}, 240901 (2018).
%
%
\bibitem{yu18} X.Z. Yu, W. Koshibae, Y. Tokunaga, K. Shibata, Y. Taguchi, N. Nagaosa, and Y. Tokura, Nature {\bf 564}, 96 (2018).
%
%
\bibitem{zhan16} X.X. Zhang,  A.S. Mishchenko, G.D. Filippis, and N. Nagaosa, Phys. Rev. B {\bf 94}, 174428 (2016).
%
%
\bibitem{fuji19} Y. Fujishiro, N. Kanazawa, T. Nakajima, X.Z. Yu, K. Ohishi, K. Kawamura, K. Kakurai, T. Arima, H. Mitamura, A. Miyake, K. Akiba, M. Tokunaga, A. Matsuo, K. Kindo, T. Koretsune, R. Arita, and Y. Tokura, Nature Comm. {\bf 10}, 1059 (2019).
%
%
\bibitem{dzya58} I. Dzyaloshinskii, J. Phys. Chem. Solid, {\bf 4}, 241 (1958).
%
%
\bibitem{mori60} T. Moriya, Phys. Rev. {\bf 120}, 91 (1960).
%
%
\bibitem{bak80} P. Bak and M.H. Jensen, J. Phys. C {\bf 13}, L881 (1980). 
%
%
\bibitem{yi09} S.D. Yi, S. Onoda, N. Nagasawa, and J.H. Han, Phys. Rev. B. {\bf 80}, 054416 (2009).
%
%
\bibitem{yu10} X.Z. Yu, Y. Onose, N. Kanazawa, J.H. Park, J.H. Han, Y. Matsui, N. Nagaosa, and Y. Tokura, Nature {\bf 465}, 901 (2010). 
%
%
\bibitem{okubo12} T. Okubo, S. Chung, and H. Kawamura, Phys. Rev. Lett. {\bf 108}, 017206 (2012).
%
%
\bibitem{lin16} S. Z. Lin and S. Hayami, Phys. Rev. B {\bf 93}, 064430 (2016).
%
%
\bibitem{hub63} J. Hubbard, Proc. R. Soc. London, Ser. A {\bf 276}, 238 (1963).
%
%
\bibitem{hub64} J. Hubbard, Proc. R. Soc. London, Ser. A {\bf 277}, 237 (1964);  {\bf 281}, 401 (1964).
%
%
\bibitem{gutz63} M.C. Gutzwiller, Phys. Rev. Lett. {\bf 10}, 159 (1963).
%
%
\bibitem{gutz64} M.C. Gutzwiller, Phys. Rev. {\bf 134}, A923 (1964); {\bf 137}, A1726 (1965).
%
%
\bibitem{kana63} J. Kanamori, Prog. Theor. Phys. {\bf 30}, 275 (1963).
%
%
\bibitem{moriya85} T. Moriya, {\it Spin Fluctuations in Itinerant Electron Magnetism} (Springer, Berlin, Heidelberg, 1985). 
%
%
\bibitem{kake13} Y. Kakehashi, {\it Modern Theory of Magnetism in Metals and Alloys} (Springer, Berlin, 2013). 
%
%
\bibitem{kake18} Y. Kakehashi, D. Koja, T. Olonbayar, and H. Miyagi, J. Phys. Soc. Jpn. {\bf 87}, 094712 (2018).
%
%
\bibitem{hay75} R. Haydock, V. Heine, and M.J. Kelley, J. Phys. C {\bf 8}, 591 (1975).
%
%
\bibitem{heine80} V. Heine, R. Haydock, and M.J. Kelley, Solid State Phys. {\bf 35}, 1 (1980).
%
%
\bibitem{merm79} N.D. Mermin, Rev. Mod. Phys. {\bf 51}, 591 (1979).
%
\bibitem{naga13} N. Nagaosa, Y. Tokura, Nat. Nanotechnol. {\bf 8}, 899 (2013).  Note that a factor of $1/4\pi$ is missing in their expression of skyrmion number $N_{sk}$ between Eqs. (B2) and (B3) in page 901.
%
%
\bibitem{igos15} P.A. Igoshev, M.A. Timirgazin, V.F. Gilmutdinov, A.K. Arzhnikov, and V.Y. Irkhin, J. Phys: Condensed Matter {\bf 27}, 446002 (2015).
%
%
\bibitem{diep13} H.T. Diep, {\it Frustrated Spin Systems} (World Scientific Pub., Singapore, 2013).
%
%
\bibitem{iked03}  T. Ikeda and Y. Tsunoda, J. Phys. Soc. Jpn. {\bf 72}, 2614 (2003).
%
%
\bibitem{kake98} Y. Kakehashi, S. Akbar, and N. Kimura, Phys. Rev. B. {\bf 14}, 8354 (1998).  
%
%
\bibitem{kimu00} N. Kimura and Y. Kakehashi, Found. Phys. {\bf 30}, 2079 (2000). 
%
%
\bibitem{uchida16} T. Uchida, Y. Kakehashi, and N. Kimura, J. Magn. Magn. Mater. {\bf 400}, 103 (2016).
%
%
 

\end{thebibliography}
\end{document}